%% file: paper.tex
\documentclass[12pt, letterpaper]{JHEP3}

\usepackage{amssymb, amsmath, amsopn, amsthm}

\usepackage{epsfig}

\usepackage{epsfig,multicol}

\usepackage{amsmath}
\usepackage{graphicx}
\usepackage{latexsym}
\usepackage{amsfonts}
\usepackage{amssymb}

\input texpref.tex

\def\makeatletter{\catcode`\@=11}
\makeatletter
\def\mathbox#1{\hbox{$\m@th#1$}}%
\def\math@ccstyles#1#2#3#4#5#6#7{{\leavevmode
     \setbox0\mathbox{#6#7}%
     \setbox2\mathbox{#4#5}%
     \dimen@ #3%
     \baselineskip\z@\lineskiplimit#1\lineskip\z@
     \vbox{\ialign{##\crcr
            \hfil \kern #2\box2 \hfil\crcr
            \noalign{\kern\dimen@}%
            \hfil\box0\hfil\crcr}}}}
\def\mathaccstyles{\math@ccstyles\maxdimen}
\def\maththroughstyles{\math@ccstyles{-\maxdimen}}
\def\unity%
{\maththroughstyles{.45\ht0}\z@\displaystyle {\mathchar"006C}\displaystyle 1}

\newcommand{\be}{\begin{eqnarray}}
\newcommand{\bea}{\begin{eqnarray}}
\newcommand{\ee}{\end{eqnarray}}
\newcommand{\eea}{\end{eqnarray}}

\title{Colored BPS Pyramid Partition Functions, Quivers and Cluster Transformations}

\author{Richard Eager$^{1}$ and Sebasti\'an Franco$^2$

\\

\vspace{0.6cm}
$^1$Institute for the Physics and Mathematics of the Universe, University of Tokyo, \\
Kashiwa, Chiba 277-8582, Japan \\
\vspace{0.3cm}
$^2$ Institute for Particle Physics Phenomenology, Department of Physics, \\
Durham University, Durham DH1 3LE, United Kingdom
\vspace{0.4cm}

\email{richard.eager@ipmu.jp, sebastian.franco@durham.ac.uk}\\

}

\abstract{
We investigate the connections between flavored quivers, dimer models, and BPS pyramids for generic toric Calabi-Yau threefolds from various perspectives.
We introduce a purely field theoretic definition of both finite and infinite pyramids in terms of quivers with flavors.  These pyramids are associated to the counting of BPS invariants for generic toric Calabi-Yau threefolds. We discuss how cluster transformations provide an efficient recursive method for computing pyramid partition functions and show that the recursion is equivalent to the multidimensional octahedron recurrence. Transitions between different pyramids are related to Seiberg dualities, and we offer complimentary characterizations of these transitions in terms of the motion of zonotopes and duality webs. 
Our methods apply to completely general geometries including those with vanishing 4-cycles, which are associated to chiral quivers, thus overcoming one of the main limitations in the existing literature. We illustrate our ideas with explicit results for the infinite family of $L^{a,b,c}$ geometries, $dP_2$, pseudo-$dP_2$, and $dP_3$. The counting of pyramid partitions for $dP_1$ gives rise to the Somos-4 sequence, while $dP_2$ and pseudo-$dP_2$ generate the Somos-5 sequence. 
Our results for $dP_3$ reproduce and extend those previously obtained for this theory, which were originally obtained from dimer shuffling.}

\preprint{IPMU 11-0197 \\ IPPP/11/79 \\ DCPT/11/158 \\ NSF-KITP-11-257}

\begin{document}

\tableofcontents

\section{Introduction}
Pyramid partitions are melting crystal configurations that correspond to a discretization of toric Calabi-Yau threefolds \cite{Okounkov:2003sp}.  They are in one-to-one correspondence with BPS states of D-branes wrapping cycles of the Calabi-Yau. 
The BPS spectrum jumps discretely at {\it walls of marginal stability} and remains locally constant inside the {\it chambers} between them. The associated pyramids are determined by both the Calabi-Yau geometry and the specific chamber under consideration. Remarkably, pyramid partition functions transform as the variables of a cluster algebra with coefficients \cite{MR2295199, MR2470108}.  The transformation properties of the partition functions follow from the work of Kontsevich and Soibelman \cite{ks} and were further explored in \cite{MR2629987, Plamondon, nagaoCluster}. 

The pyramids under consideration are intimately related to periodic quivers and brane tilings. Stones in the pyramids correspond to certain paths on quivers and the pyramid partitions obtained by removing some of these stones are in one-to-one correspondence with perfect matchings on brane tilings.
While there is an extensive literature on this topic, it mainly applies to toric Calabi-Yau threefolds without vanishing 4-cycles.
This condition places a severe constraint on the associated quiver gauge theories, restricting them to being non-chiral. The study of pyramid partitions and wall crossing for chiral quivers has been initiated in \cite{arXiv:1006.2113}. The main goal of this paper is to initiate a systematic study of the pyramids associated to general brane tilings, i.e. including those associated to chiral quivers, combining tools from gauge theory, dimer models, toric geometry, and cluster algebras.

This paper is organized as follows. In Section \ref{section_quivers_tilings_geometry}, we review various concepts related to quiver gauge theories, brane tilings and toric geometry. Section \ref{section_quivers_and_geometry} discusses the connection between quivers and geometry from various perspectives. Here we discuss the concept of a zonotope, which later becomes useful for analyzing the space of Seiberg dual theories. In Section \ref{section_pyramids_from_quivers}, we introduce a field theoretic definition of pyramids of both infinite and finite type in terms of framing flavors. This definition applies to arbitrary toric geometries, including those with compact 4-cycles, i.e. those giving rise to chiral quivers. Given a field theory definition of pyramids, it is possible to investigate how they transform under Seiberg duality. We explain this point in Section \ref{section_pyramids_Seiberg_duality}, where we see that Seiberg duality changes the type and number of top stones. Pyramid partitions and their connection to BPS invariants is the subject of Section \ref{section_pyramid_partitions}. Section \ref{section_recursive} explains how to use cluster mutations to recursively generate pyramid partition functions. This procedure is highly efficient and generates complicated partition functions from trivial initial data, without needing to explicitly construct the pyramids. In this section we also discuss a physical perspective on cluster transformations in terms of quivers with flavors. We introduce a recursive procedure for constructing the shadow of a pyramid and we show cluster transformations can be casted as the multidimensional octahedron recurrence. Generalizing what happens for simple geometries, it is natural to expect that Seiberg duality is responsible for transitions between stability chambers. Section \ref{section_duality_cascades} studies the space of Seiberg dual theories from the complementary viewpoints of geometry, zonotopes, and duality webs.
Section \ref{section_explicit_examples} contains explicit examples: the infinite $L^{a,b,c}$ family of geometries, $dP_3$, $dP_2$, and $PdP_2$. We also provide a change of variables from quiver gauge groups to fractional brane charges that leads to convergent expressions for partition functions as the number of mutations goes to infinity. Appendix \ref{section_dP2_and_dP3} collects further details on some of the models analyzed in this section. 
We conclude in Section \ref{section_conclusion}.

\bigskip
\bigskip

\section{Quivers, Brane Tilings, and Geometry}

\label{section_quivers_tilings_geometry}

In this section we compile some background material that will be used throughout the paper.

\subsection{Quiver Gauge Theories and Brane Tilings}

We consider gauge theories that describe the low energy physics on the world-volume of a stack of coincident D3-branes placed at a Calabi-Yau singularity. The gauge theories obtained in this manner can be succinctly described in terms of a quiver.  A quiver $Q = (V, A, h, t)$ is a collection of
vertices $V$ and arrows $A$ between the vertices of the quiver.  The maps $h$ and $t$ define the head and tail of a given arrow. The vertices or nodes of the quiver represent gauge groups, i.e. vector multiplets, and the arrows represent bifundamental or adjoint chiral multiples.

For the special case of toric Calabi-Yau singularities, the corresponding quiver gauge theory has a simple graphical description in terms of a brane tiling \cite{Franco:2005rj}. A {\it brane tiling} is a bipartite graph $G = (G_0^{\pm}, G_1)$ embedded into the two-torus such that the faces form a tiling of the torus.
A {\it periodic quiver} $Q = (Q_0, Q_1, Q_2,h,t)$ is obtained from the dual graph of the brane tiling.  The vertices $Q_0$ of the periodic quiver are dual to the faces of the brane tiling.  The plaquettes $Q_2 = Q_2^{+} \cup Q_2^{-}$ of the  periodic quiver have clockwise and counterclockwise orientation respectively, where the orientation is determined from the bipartite structure of the dual graph.  This allows us to define the superpotential $W \in \mathbb{C} Q / [\mathbb{C} Q, \mathbb{C} Q]$ as the sum over all plaquettes with the sign of each term given by the corresponding orientation,

\beq
W = \sum_{P \in Q_2^{+}} w_P - \sum_{P \in Q_2^{-}} w_P,
\eeq
where the word $w_P$ is defined to be the products of all of the arrows around the plaquette $P$.  The {\it superpotential algebra} $\mathcal{A} = \mathbb{C} Q / (\partial W)$ is obtained from identifying elements in the path algebra using the relations given by the partial derivatives of the superpotential.

A {\it representation} $X$ of a quiver $Q$ with {\it dimension vector} $\mathbf{n} \in \mathbb{N}^{|V|}$ is a collection of vector spaces $X_v$ of dimension $n_v$ and maps $\phi_a : X_{t(a)} \rightarrow X_{h(a)}$ corresponding to the vertices, $v,$ and arrows, $a,$ of the quiver.
In this paper we will be concerned with quiver gauge theories with gauge group 
\beq
G = \prod_{v \in V} U(n_v).
\eeq
Each factor in the gauge group is a unitary group with size determined by the dimension vector of a quiver representation.
The moduli space of vacua of the 4D $\mathcal{N} = 1$ supersymmetric field theory is naturally encoded by representations of the superpotential algebra $\mathcal{A} = \mathbb{C} Q / (\partial W).$

\begin{figure}[h]
\begin{center}
\includegraphics[width=13cm]{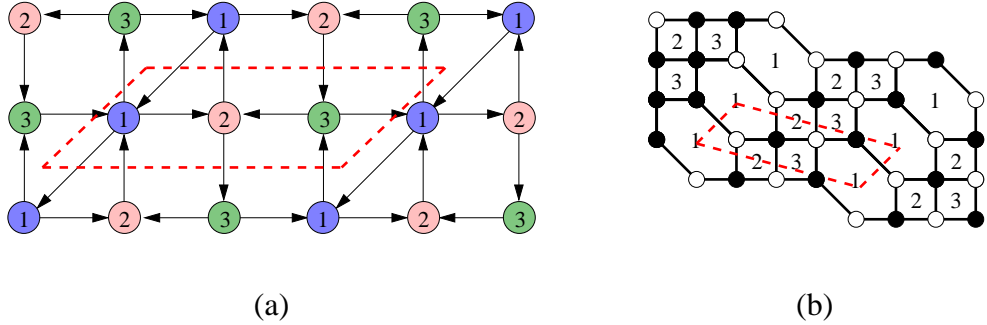}
\caption{a) Periodic quiver and b) brane tiling for the SPP. We indicated the corresponding unit cells with dashed red lines.}
\label{quiver_dimer_SPP}
\end{center}
\end{figure}

In the sections that follow, we will use the example of the Suspended Pinch Point (SPP) to illustrate various ideas. \fref{quiver_dimer_SPP} shows the periodic quiver and brane tiling for the SPP.  

\subsection{Toric Geometry}

\label{section_toric_geometry}

Here we quickly review the basics of toric geometry. We refer the reader to \cite{MR1234037, MR2810322} for a more detailed presentation.
A toric variety can be constructed from a lattice $N \cong \mathbb{Z}^n$ and a fan $\mathcal{F}.$  A fan $\mathcal{F}$ is a collection of convex rational polyhedral cones in
$N_{\R} := N \otimes_{\mathbb{Z}} \R$ satisfying additional incidence relations which can be found in \cite{MR1234037, MR2810322}.
Given a convex polyhedral cone $\sigma \in N_{\R}$, the dual cone $\sigma^{\vee}$ is the set of vectors in the dual lattice $M = \Hom(N, \mathbb{Z})$
 that are nonnegative on $\sigma$.  This collection of vectors forms a commutative semigroup
 $$\mathcal{S}_{\sigma} = \sigma^{\vee} \cap M = \{ m \in M \; \vert \; \langle m,v \rangle \ge 0 \text{ for all } v \in \sigma \}.$$
Given a rational polyhedral cone $\sigma = [v_1, v_2, \dots, v_D]$ we define the ring $R_{\sigma} := \C[\sigma^{\vee} \cap M]$
and the variety $X_{\sigma} = {\rm Spec} \, R_{\sigma}.$ For toric Calabi-Yaus, the first coordinate of all $v_i$ can always be set to $1$.
This means that in the case of complex dimension $n=3$ we can represent any toric Calabi-Yau cone by a convex lattice polytope in $\mathbb{Z}^2$, where the positions of vertices are given by the remaining two coordinates.
This polytope is usually called the toric diagram.

\bigskip

\section{From Quivers to Geometry and Back}

\label{section_quivers_and_geometry}

One of the main successes of brane tilings is that they completely trivialized the computation of the Calabi-Yau geometry probed by D3-branes, which corresponds to the moduli space of the quiver gauge theories on the D3-branes. The main ingredient in this simplification is a one-to-one correspondence between GLSM fields in the toric description of the Calabi-Yau (i.e. points in the toric diagram) and perfect matchings \cite{Franco:2005rj}. Perfect matchings and their positions in the toric diagram can be immediately determined by computing the determinant of the Kasteleyn matrix, an adjacency matrix of the brane tiling. Methods for going in the opposite direction, i.e. for determining a brane tiling, equivalently a quiver gauge theory, starting from a toric Calabi-Yau threefold have been introduced in \cite{Hanany:2005ss,hep-th/0511287}. 
Here we present a new approach for connecting geometry and brane tilings, which will be useful for the discussion in Section \ref{section_duality_cascades}.

\subsection{Geometry from Periodic Quivers}

\label{section_R_charge}

Given a brane tiling with an R-charge, we will explain how to reconstruct the Calabi-Yau geometry. In the process we will introduce a set of coordinates based on $R$-charge assignments, which we denote $\Psi$-coordinates, for the gauge groups in the quiver. The discussion in this section is closely related to the ideas introduced in \cite{Butti:2005vn}. A {\it consistent R-charge} is a charge assignment to the chiral matter fields of the quiver, $R: Q^1 \rightarrow \mathbb{R}$, satisfying two constraints to ensure that the resulting theory is superconformal. These constraints are simply that the NSVZ beta functions for all gauge groups and the beta functions for all superpotential couplings vanish. For toric quiver theories, these constraints simplify to the following geometric conditions \cite{Franco:2005rj} 
\beq
\begin{array}{rcccl}
\sum_{a \in P} R(a) & = & 2 & \qquad & \text{for all plaquettes } P \in Q_2 \\
\sum_{a \in V} (1 - R(a)) & = & 2 & \qquad & \text {for all quiver nodes } V \in Q_0
\end{array}
\eeq
where $a$, $P$ and $V$ indicate arrows, plaquettes, and nodes in the periodic quiver, respectively.
Alternatively, these constraints can be interpreted as sums over faces and nodes of the brane tiling dual to the periodic quiver \cite{Franco:2005rj}.
The space of all R-charge assignments is convex linear combination of R-charges associated to every point on the boundary of the toric diagram, which we denote $B$.
\footnote{
We allow the possibility of having internal points in $B$, i.e. of having three or more collinear points in $B$.
This case corresponds to non-isolated singularities. In general, more than one perfect matching can be associated to internal points in $B$. The way perfect matchings contribute to chiral fields is that all perfect matchings that correspond to the same point in the toric diagram appear simultaneously, raised to the same power. This means that, even in this situation, we can effectively consider a single contribution from every point in $B$, and our discussion applies without modifications. \label{internal_points_in_B}}

Let us associate a vector $E_i=(0,\ldots,0,1,0,\ldots,0)$, with $1$ in the $i^{th}$ entry, $i=1,\ldots,D$, to every perfect matching in $B$.
For each edge $e$ of the brane tiling we assign the vector
\beq
\Psi(e) = \sum_{i \in B \text{ such that } e \in p_i} E_i \qquad
\label{psi_e}
\eeq
in $\mathbb{Z}^D,$ where the sum is over the set of boundary perfect matchings the edge belongs to. 
We also assign this same vector to the corresponding dual arrow in the periodic quiver.
Following the discussion in footnote \ref{internal_points_in_B}, we should add a single contribution for every point in $B$, even when multiple perfect matchings might correspond to the same point. 
The vector associated to a path is defined using linearity to be the sum of the vectors associated to arrows in the path,
$\Psi(\gamma) = \sum_{e \in \gamma} \Psi(e).$
For any plaquette $P$ of the periodic quiver $\Psi(P) = \sum_{i} E_i.$\footnote{This is a straightforward consequence of the well-known fact that every superpotential term in these quivers contains all perfect matchings, which implies that it contains all boundary perfect matchings.}
Any consistent $R$-charge assignment can be written as a linear map $\hat{R}: \mathbb{Z}^D \rightarrow \mathbb{R}$ subject to the single constraint $\hat{R}(\sum_{i} E_i) = 2,$ i.e. that the total $R$-charge of the boundary points in the toric diagram is equal to 2 \cite{Butti:2005vn}. The $R$-charge of any path $\gamma$ is given by the composite map
\beq
R(\gamma) = \hat{R} \circ \Psi(\gamma).
\eeq
Conversely any such map satisfying the constraint defines a consistent $R$-charge assignment.  
Any two paths $\gamma$ and $\gamma'$ in the same homology class of the torus have the same image under $\Psi$ up to an integral multiple of $\sum_{i} E_i,$
\beq
\Psi(\gamma) = \Psi(\gamma') + m \sum_{i} E_i, \qquad \text{for some } m \in \mathbb{Z}.
\eeq

Given a fundamental domain for the periodic quiver we choose two paths $\alpha$ and $\beta$ that span the two homology classes of the torus.
The paths are far from unique, but the corresponding cycles are uniquely defined up to the action of $SL(2,\mathbb{Z}).$  
The boundary perfect matching content $\Psi(\alpha)$ and $\Psi(\beta)$ allows us to reconstruct the toric diagram as follows.  For each point in the boundary of the toric diagram $i \in B$ there is a corresponding vector
$v_i = (\Psi(\alpha)_i, \Psi(\beta)_i).$  These vectors are the 2D coordinates of the points in the toric diagram.  
The vectors $v_i$ in our construction are only defined up to the action of $SL(2,\mathbb{Z}),$ but toric diagrams differing by $SL(2,\mathbb{Z})$ are equivalent in toric geometry.
Since the points form a convex $D$-gon, they have a natural cyclic order. The $\Psi$ map we have just defined is similar to the one introduced in \cite{Hanany:2006nm}.

\subsection*{Example}

As an example, we show the periodic quiver for $dP_2$ in \fref{fig:dP2tiling}, where we indicate the two paths $\alpha$ and $\beta$ spanning the homology of the torus. We arrange $\Psi(\alpha)$ and $\Psi(\beta)$ as the rows of the matrix
\beq
\begin{pmatrix}
 \ 0 \ & \ 1 \ & \ 2 \ & \ 1 \ & \ 0 \ \\
 -1 & -1 & 0 & 1 & 0
\end{pmatrix}.
\label{matrix_toric_dP2}
\eeq
The column vectors are precisely the vertices of the toric diagram for $dP_2$, which is shown in \fref{fig:toric_dP2}. 

\subsection{Periodic Quivers from Geometry}

In the previous section we discussed one approach for determining the geometry associated to a periodic quiver. We now explain the reverse procedure of constructing a periodic quiver from geometry. Methods for achieving this goal were originally introduced in \cite{Hanany:2005ss,hep-th/0511287}.
However this new construction is particularly well suited for describing periodic duality cascades.

We first recall the relationship between tilting objects and periodic quivers.  We next explain how to find a tilting object using a zonotope constructed from toric data.  We then complete the discussion by explaining how to obtain a periodic quiver from a tilting object.

\subsubsection{Tilting Objects and Periodic Quivers}
\label{sec:tilting}

We now explain how the $\Psi$-coordinates of each node in the quiver have a natural interpretation in terms of modules.

Following the discussion in Section \ref{section_toric_geometry}, gauge invariant operators are represented by closed paths in the quiver and correspond to the elements $m$ in
\beq
\mathcal{S}_{\sigma} = \sigma^{\vee} \cap M = \{ m \in M \; \vert \; \langle m,v \rangle \ge 0 \text{ for all } v \in \sigma \},
\eeq
where $[v_1, v_2, \dots, v_D]$ correspond to vertices of the toric diagram. This allows us to define the ring $R_{\sigma} := \C[\sigma^{\vee} \cap M]$ which corresponds to the singular geometry.  
Our goal in this section is to explain how each gauge group in the periodic quiver corresponds to a module over the ring $R_{\sigma}.$

We first give a description of modules directly in terms of the fan.  Given a $D$-tuple of integers $(b_1, \dots b_D)$ we define the semigroup module
$\mathbb{T}(b)$ over the semigroup ring $\mathbb{T}(0) = \sigma^{\vee} \cap M$ by 
\beq
\mathbb{T}(b) := \{m \in M | \langle m,v_i, \rangle \ge b_i \}
\eeq
and define the module $T(b) := \rm{Span}_{\C} \mathbb{T}(b).$
Two $R$-modules $T(b)$ and $T(b')$ are isomorphic if and only if there exists an $m \in M$ such that
$b_i = b_i' + \langle m, v_i \rangle$ \cite{Bocklandt:2011vv, Per05}.

We will assign modules to every node in the quiver using the $\Psi$ map as follows.  First we must extend the $\Psi$ map from paths to nodes.
Fix any node $n_0$ of the quiver to have coordinate $\Psi(n_0) = \vec{0} \in \mathbb{Z}^D.$  Then all other vertices $n$ can be given by paths $\gamma_n$ from $n_0$ to $n.$  While the path is not uniquely defined, we will see that the corresponding module
$T \circ \Psi(\gamma_n)$ is uniquely defined.  The image $\Psi(\gamma_n)$ is only defined up to an integral linear combination of $\Psi(\alpha), \Psi(\beta),$ and $\Psi(P).$
If we form a matrix with these three vectors as its rows, then the columns of the matrix
\beq
\begin{pmatrix}
\Psi(\alpha) \\
\Psi(\beta) \\
\Psi(P)
\end{pmatrix}
\eeq
are the vectors $v_i$, $i=1,\ldots,D$ of the toric diagram.
This is precisely the isomorphism between modules $T(b)$ and $T(b')$ with $b_i = b_i' + \langle m, v_i \rangle.$

Returning to our example of $dP_2,$ for our choice of the fundamental domain we see that the nodes of the quiver correspond to the modules $T(b_\alpha)$ given by
\beq
T(0,0,0,0,0), T(1,0,0,0,0),T(1,1,0,0,0),T(1,1,1,0,0),T(1,1,1,1,0),
\eeq
where $\alpha = 1, 2, \dots , 5$ labels the five nodes.
The superpotential algebra $A \cong \mathbb{C} Q / (\partial W)$ is then
\beq
A \cong \rm{End} \left( \bigoplus T(b_\alpha) \right).
\eeq
Below we explain this isomorphism and how to recover the quiver directly from the tilting object.

\subsubsection{Constructing a Tilting Object from Geometry: Zonotopes}
\label{sec:zonopq}

We have just discussed the connection between a tilting object and gauge groups in a quiver. As a first step in the determination of the quiver we now explain how to construct a tilting object starting from the geometry of a toric Calabi-Yau threefold. The procedure described in this section is an adaptation of the construction of a tilting bundle on a Fano stack \cite{MR2509327} to local Calabi-Yau singularities.

In what follows, we restrict to the case in which the boundary of the toric diagram does not contain internal points for simplicity. The construction we discuss can be extended to this case. Consider the lattice $\mathbb{Z}^D$ with basis vectors $E_i.$  In the previous section we identified modules $T(b)$ and $T(b')$ as being isomorphic if their weights satisfied
\beq
b_i = b_i' + \langle m, v_i \rangle.
\eeq
We quotient the lattice $\mathbb{Z}^D$ by this equivalence relation, and call the images of the basis vectors $\widehat{E}_i.$
Call the common image of $b$ and $b'$ under the equivalence relation $\widehat{b}.$
Each lattice point $\widehat{b}$ in $\mathbb{Z}^{D-3}$ determines a module $T(\widehat{b}).$
Arranging these $D$ vectors as the columns of a $D$ by $D-3$ matrix, we see that the rows have a simple interpretation as GLSM charge vectors.
We construct the polytope 
$$Q = \lambda_1 \widehat{E}_1 + \lambda_2 \widehat{E}_2 + \dots + \lambda_D \widehat{E}_D, \qquad \lambda_1, \lambda_2, \dots \lambda_D \in [0,1].$$
Since $Q$ is the Minkowski sum of intervals $ [0, \widehat{E}_j],$ it is a zonotope. 
One of the key properties of $Q$ is that its faces are in one-to-one correspondence with intersecting diagonals in the toric diagram.

Next we construct a polytope $\widehat{P}$ from $Q.$  The interior lattice points of $\frac{1}{2} \widehat{P}$ will correspond to the gauge groups of a quiver gauge theory.
The defining property of $\widehat{P}$ is that it is a centrally symmetric polytope such that the midpoints of all of its faces are the vertices of $Q.$  Additionally each vertex of $Q$
must be the midpoint of some face of $\widehat{P}.$  The construction of $\widehat{P}$ is not unique, however the additional data is given by a choice of R-charge.  

Each integral lattice point $\widehat{b}$ inside the zonotope $\frac{1}{2} \widehat{P}$ uniquely specifies a module $T(\widehat{b}).$  
The direct sum of these modules forms the tilting module
$$\mathcal{S} = \bigoplus_{\widehat{b} \in \frac{1}{2} \widehat{P}} T(\widehat{b}).$$
The superpotential algebra is the endomorphism algebra of the tilting object, $\mathcal{A} = \rm{End}_R(\mathcal{S}).$ 
In the next section we explain how to complete the determination of the gauge theory, i.e how to add the matter content and superpotential to the already identified gauge groups.

\bigskip

\subsubsection*{Examples}
\begin{figure}[h]
\centering
\begin{tabular}{ccc}
\epsfig{file=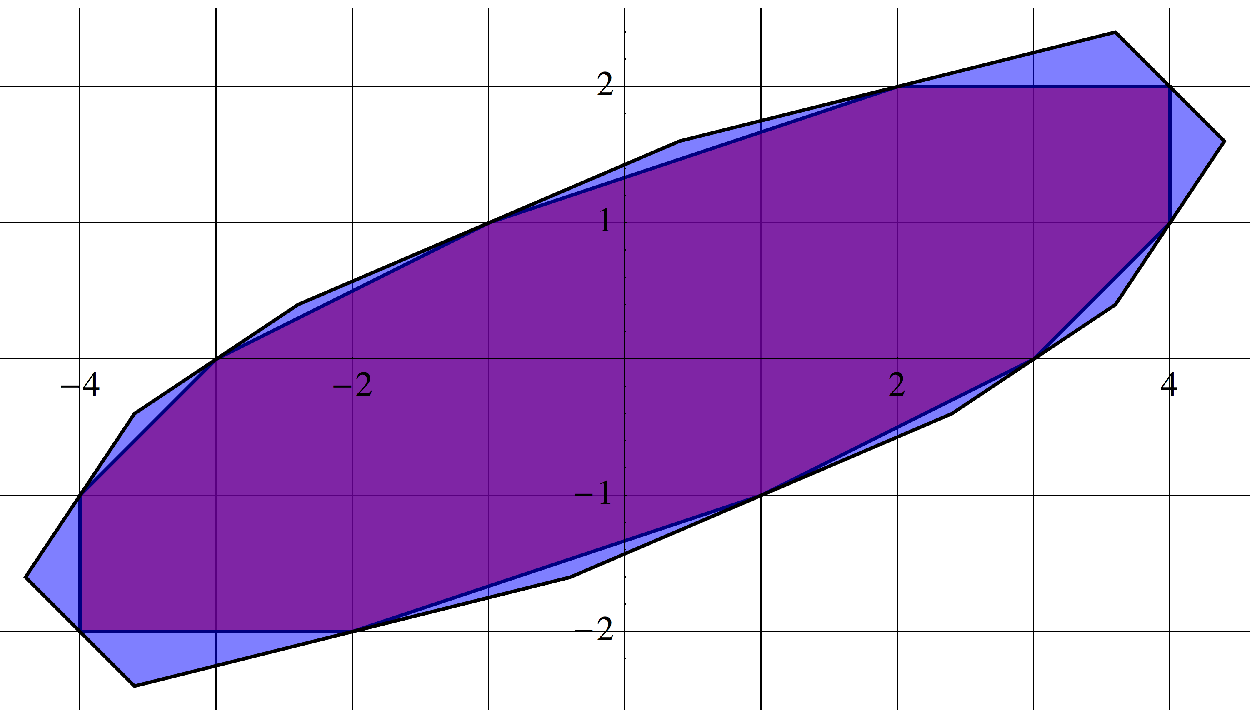,width=0.4\linewidth,clip=} & \ \ \ &
\epsfig{file=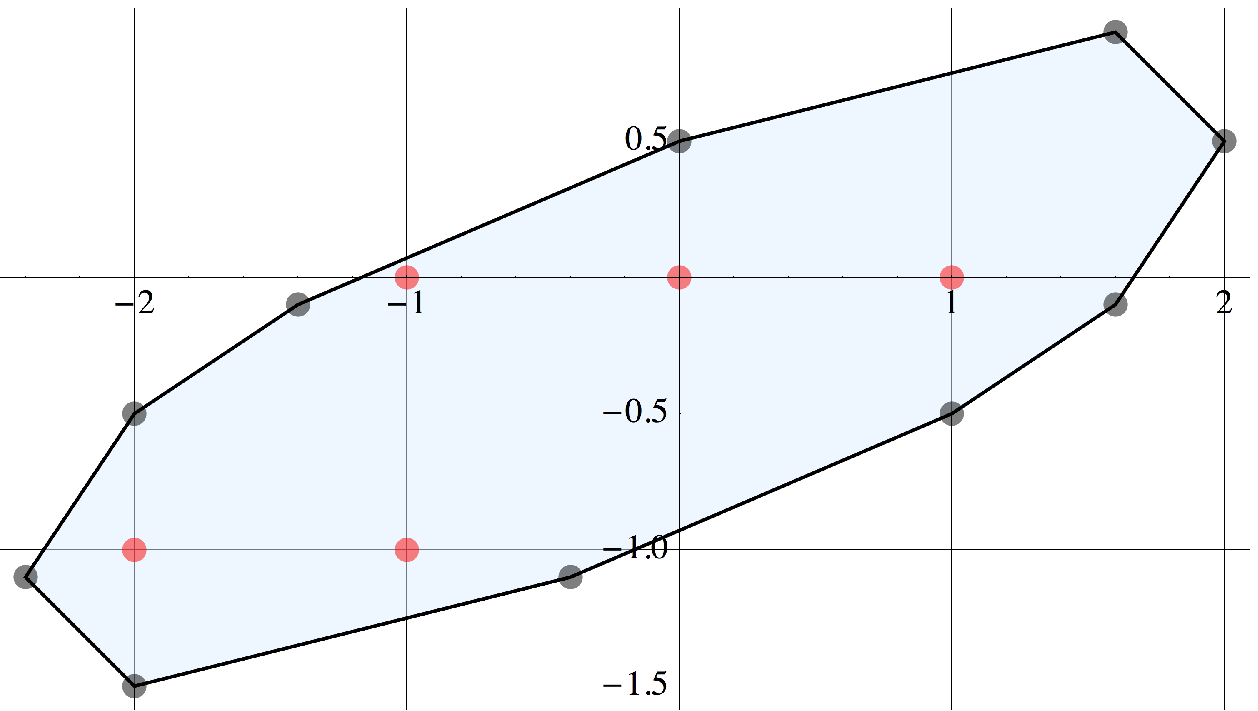,width=0.4\linewidth,clip=} \\
(a) & & (b)
\end{tabular}
\caption{a) Zonotopes $Q$ (purple) and $\widehat{P}$ (blue). b) Zonotope $\frac{1}{2} \widehat{P}.$ The five interior points, corresponding to the five gauge groups in the $dP_2$ quiver, are shown in red.}
\label{zonotope_dP2}
\end{figure}

\begin{figure}[h]
\begin{center}
\includegraphics[width=5.4cm]{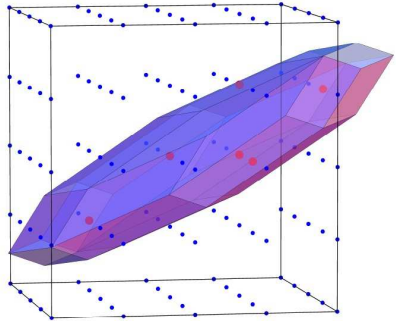}
\caption{Zonotope $\frac{1}{2} \widehat{P}$ for $dP_3$. The six internal points, corresponding to the six gauge groups in the $dP_3$ quiver, are shown in red.}
\label{dP3zonoX}
\end{center}
\end{figure}

Here we illustrate the previous construction with $dP_2$ and $dP_3$ as examples.
\begin{figure}[h]
\begin{center}
\includegraphics[width=2.7cm]{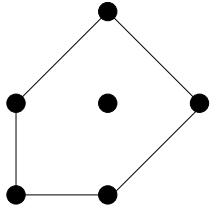}
\caption{Toric diagram for $dP_2$.}
\label{fig:toric_dP2}
\end{center}
\end{figure}
The toric diagram for $dP2$ is shown \fref{fig:toric_dP2} with vertices at
\beq
\begin{pmatrix}
 \ 0 \ & \ 1 \ & \ 2 \ & \ 1 \ & \ 0 \ \\
 -1 & -1 & 0 & 1 & 0
\end{pmatrix}.
\label{matrix_GLSM_charges_dP2}
\eeq
The vectors in the toric diagram satisfy the following two linear relations
\beq
\begin{pmatrix}
 2 & -3 & 2 & -1 & 0 \\
 \ 0 \ & -1 & \ 1 \ & -1 & \ 1 \ 
\end{pmatrix}.
\label{matrix_toric_dP2}
\eeq
The rows in \eref{matrix_toric_dP2} are obtained by the condition of orthogonality to the rows of \eref{matrix_GLSM_charges_dP2} and the $(1,1,1,1,1)$ vector, up to $SL(2,\mathbb{Z})$ transformations. The column vectors of the matrix of relations are denoted by $\widehat{x}_j$ where $j \in [1, \dots, D],$ and will be used in the next section to construct the arrows of the periodic quiver. The blue shape in \fref{zonotope_dP2}.a is the zonotope $\widehat{P}.$ 
Figure \ref{zonotope_dP2}.b shows the collection of interior lattice points in  $\frac{1}{2} \widehat{P}.$ 
The zonotope for $dP_3$ lives in three dimensions and has 15 pairs of parallel faces. We show it in \fref{dP3zonoX}.

\subsubsection{Constructing the Periodic Quiver from a Tilting Object}
\label{section_quiver_from_tilting}

We have just explained how to construct a tilting object, which corresponds to the gauge groups of the quiver, for any toric Calabi-Yau singularity. In this section, we show how to determine a periodic quiver, i.e. how to add to these gauge groups the matter fields and determine the superpotential, associated to the tilting object. The construction of a gauge theory from a tilting object is explained in \cite{Aspinwall:2008jk, Eager:2010ji} The reader is urged to consult these papers for definitions of unfamiliar terms. However at the end of this section we will give a simple algorithmic construction of the quiver gauge theory independent of the mathematics used in the construction.

\begin{figure}[h]
\begin{center}
\includegraphics[width=6cm]{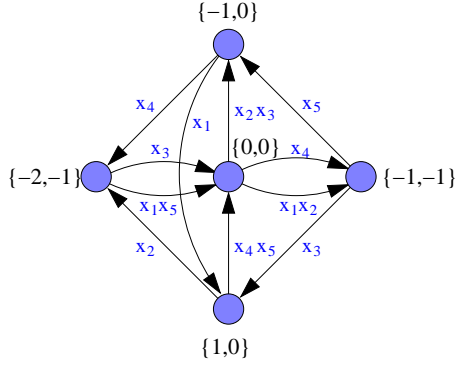}
\caption{$dP_2$ quiver constructed from the zonotope $\frac{1}{2} \widehat{P}$. We indicate the modules $T(\hat{b}_i)$ associated to each gauge group and the module homomorphisms corresponding to arrows.}
\label{fig:dP2quiver}
\end{center}
\end{figure}

The arrows in the quiver are given by the irreducible morphisms in $\rm{End}_R(\mathcal{S}).$ We will use the symbols $x_j, j = 1, \dots, D$ to represent the morphisms in $\rm{End}_R(\mathcal{S})$ induced by multiplication. 
Each of these morphisms corresponds to a column vector $\widehat{x}_j$ in the GLSM matrix. There is an arrow from vertex $\widehat{b}_1$ to vertex $\widehat{b}_2$ in the quiver if and only if
$\widehat{b}_1 + \widehat{x}_j = \widehat{b}_2.$
For any subset $\pi \in [1, \dots D]$ there is a morphism from vertex $\widehat{b}_1$ to vertex $\widehat{b}_2$ if and only if
$\widehat{b}_1 + \sum_{j \in \pi} \widehat{x}_j = \widehat{b}_2.$
However only the irreducible morphisms label arrows of the quiver.
This is the case if and only if there is no proper subset of $\pi$ that corresponds to a morphism. We illustrate this construction for $dP_2$ in \fref{fig:dP2quiver}.  For example, there is an arrow ``$x_2$'' from $T(1,0)$ to $T(-2,-1)$ since the corresponding module homomorphism is $x_2 = (-3,-1)$, the second column of the GLSM matrix \eref{matrix_toric_dP2}.

\begin{figure}[h]
\begin{center}
\includegraphics[width=10.5cm]{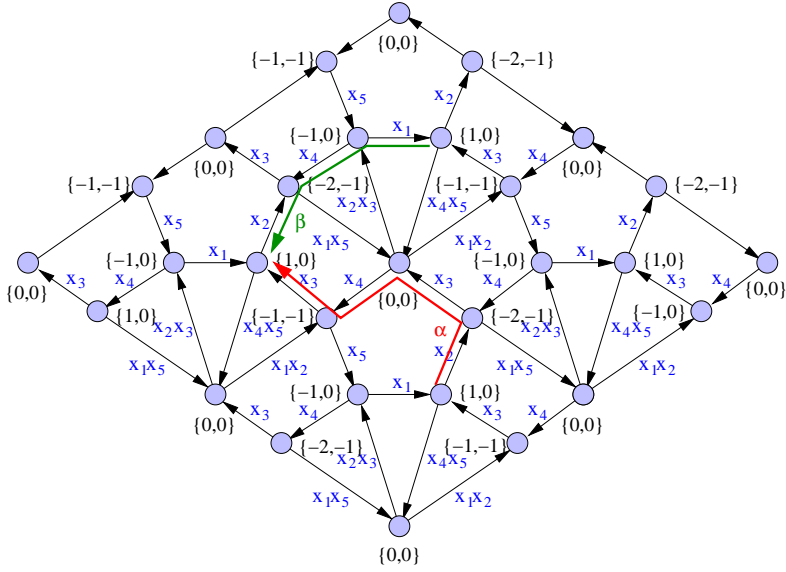}
\caption{Periodic quiver for $dP_2$. We indicate the modules $T(\hat{b}_i)$ associated to each gauge group and the module homomorphisms corresponding to arrows. We also show the choice of $\alpha$ and $\beta$ paths that results in \eref{matrix_toric_dP2}.}
\label{fig:dP2tiling}
\end{center}
\end{figure}

We can similarly construct the full periodic quiver, which in addition encodes the superpotential via its plaquettes. 
Given the elements $\widehat{b} \in \frac{1}{2} \widehat{P}$ we choose a lift to $b_1, b_2, \dots, b_{M} \in \mathbb{Z}^D.$
We choose a map $\pi: \mathbb{Z}^D \rightarrow \R^2$ such that $\pi(E_i)$ form the edges of a closed convex $D$-gon. \fref{fig:dP2tiling} shows the periodic quiver constructed this way for $dP_2$.

\section{Pyramids from Quivers}

\label{section_pyramids_from_quivers}

In this section we introduce a field theoretic definition of certain infinite and finite pyramids, which are the main focus of this paper, in terms of quivers with flavors. This construction is a generalization of the one discussed in \cite{Chuang:2008aw} for the conifold.

\subsection{Framing: Flavors and Superpotential Relations}

\begin{figure}[h]
\begin{center}
\includegraphics[width=10cm]{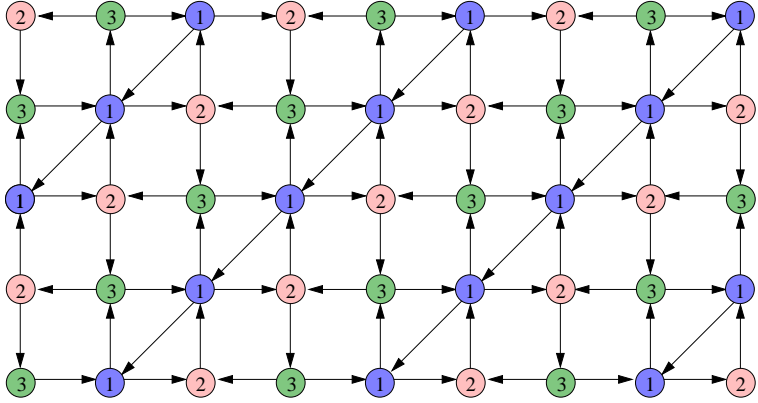}
\caption{Periodic quiver for the SPP.}
\label{periodic_quiver_SPP}
\end{center}
\end{figure}

The starting point for constructing pyramids is the periodic quiver associated with the geometry under study.
\fref{periodic_quiver_SPP} shows this object for the SPP.
Let us construct an infinite pyramid with $n$ top stones of type $\alpha_i$ ($i=1,\ldots,n$), i.e. corresponding to gauge groups $\alpha_i$ in the quiver.
From a quiver point of view, this corresponds to introducing flavors $q_i$ ($i=1,\ldots,n$) in the fundamental representation of gauge group $\alpha_i$.
Stones in the pyramid correspond to chiral operators and are given by oriented paths in the quiver that have a $q_i$ field at one of its endpoints.
Notice that all these paths are open.\footnote{It is important to emphasize that all arrows in the quiver, including those of framing flavors, can be reversed by taking a convention in which fundamental and antifundamental representations of gauge groups are switched. This orientation flip has absolutely no effect on the physics. In this equivalent convention, stones in the pyramid correspond to oriented paths starting with a $q_i$.}   

The vertical position of a stone is determined by the R-charge (equivalently the conformal dimension) of the corresponding chiral operator. The operators associated with stones that are on top of each other differ on a number of plaquettes in the periodic quiver, which correspond to superpotential terms and, as a result, their R-charges differ by a multiple of 2. 

Next, we introduce flavors $p_j$ ($j=1,\ldots,n-1$) in the antifundamental representation of the gauge groups $\beta_j$. These flavors allow us to introduce gauge invariant superpotential couplings that can be put in the general form

\beq
W_{rels}=p_1 \, \mathcal{O}_1 \, q_1+ (p_2 \, \mathcal{O}_2 + p_1 \, \tilde{\mathcal{O}}_1) q_2 +\ldots+(p_{n-1} \, \mathcal{O}_{n-1} + p_{n-2} \, \tilde{\mathcal{O}}_{n-2}) q_{n-1}+ p_{n-1} \, \tilde{\mathcal{O}}_{n-1} \, q_{n}
\label{W_framing_1_Kahler}
\eeq
The operators $\mathcal{O}_i$ and $\tilde{\mathcal{O}}_j$  ($i,j=1,\ldots,n-1$) can be read off directly from the periodic quiver. They correspond to the {\it shortest} paths connecting the pairs of flavors they couple to. Longer paths connecting the same nodes differ from the shortest ones by closed loops. 

As we have mentioned, we will define stones in the pyramid as open oriented paths containing a $q_i$ field at one of its endpoints. We can eliminate paths containing $p_j$ fields by setting $p_j=0$ for all $j$.

Next, let us consider the F-term relations coming from framing flavors. Since we have $p_j=0$ all the conditions $F_{q_i}=0$ are automatically satisfied and we are left with the vanishing of the F-terms for $p$'s, which result in $(n-1)$ relations:
\beq
\mathcal{O}_i \, q_i=\tilde{\mathcal{O}}_i \, q_{i+1} \ \ \ \ \ \ \ i=1,\ldots,{n-1}.
\label{relations_framing_with_q}
\eeq
We refer to the previous steps as {\it framing with $q$'s}. They can be summarized as follows:

\bigskip
\medskip

\noindent\hspace{.4cm}\begin{tabular}{|c|}
\hline
{\bf Framing with $q_i$: Infinite Pyramids} \\ \\
\begin{tabular}{rl}
{\bf 1)} & Introduce flavors $q_i$ transforming in the fundamental representation of \\ & gauge groups $\alpha_i$, $(i=1,\ldots, n)$. \\ & \\

{\bf 2)} & Introduce flavors $p_j$ transforming in the antifundamental representation \\ & of gauge groups $\beta_j$, $(j=1,\ldots, n-1)$. \\ & \\

{\bf 3)} & Add to the superpotential the interactions \eref{W_framing_1_Kahler}. \\ & \\
& Every stone corresponds to an open oriented path ending with a $q_i$. Paths  \\ & containing a $p_j$ are eliminated by setting $p_j=0$, which implies that $F_{q_i}=0$ \\ & and we are left with the F-term equations of the $p_j$ as relations. There are \\ & $n$ top stones and $(n-1)$ relations, resulting in an {\it infinite} pyramid.
\end{tabular} \\ 
\hline
\end{tabular}

\bigskip

\bigskip

\begin{figure}[h]
\begin{center}
\includegraphics[width=10cm]{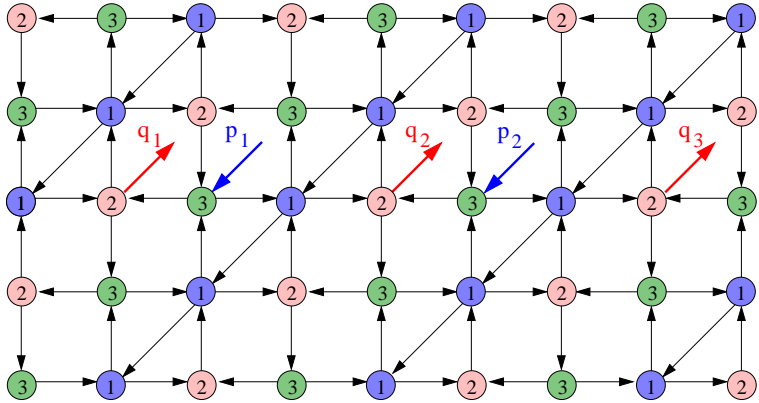}
\caption{Flavored quiver for SPP with $n=3$ top stones. $q_i$'s are indicated in red and $p_j$'s in blue. The configuration corresponds to $\mathcal{O}_i=X_{32}$ and $\tilde{\mathcal{O}}_j=X_{31} X_{12}$.}
\label{periodic_quiver_SPP_framing}
\end{center}
\end{figure}

We now illustrate these ideas with the SPP example. Let us consider $n=3$ top stones of type 2. In order to specify the relative positions of these three stones, we need to determine the quiver nodes to which we add the extra flavors.
Consider placing all $p_j$ fields at nodes to type 3. \fref{periodic_quiver_SPP_framing} illustrates this specific configuration of flavors.  From this quiver, we can immediately determine that $\mathcal{O}_i=X_{32}$ and $\tilde{\mathcal{O}}_j=X_{31} X_{12}$, $i,j=1,2$.

\subsubsection{Constructing the Pyramids}

\begin{figure}[h]
\begin{center}
\includegraphics[width=12cm]{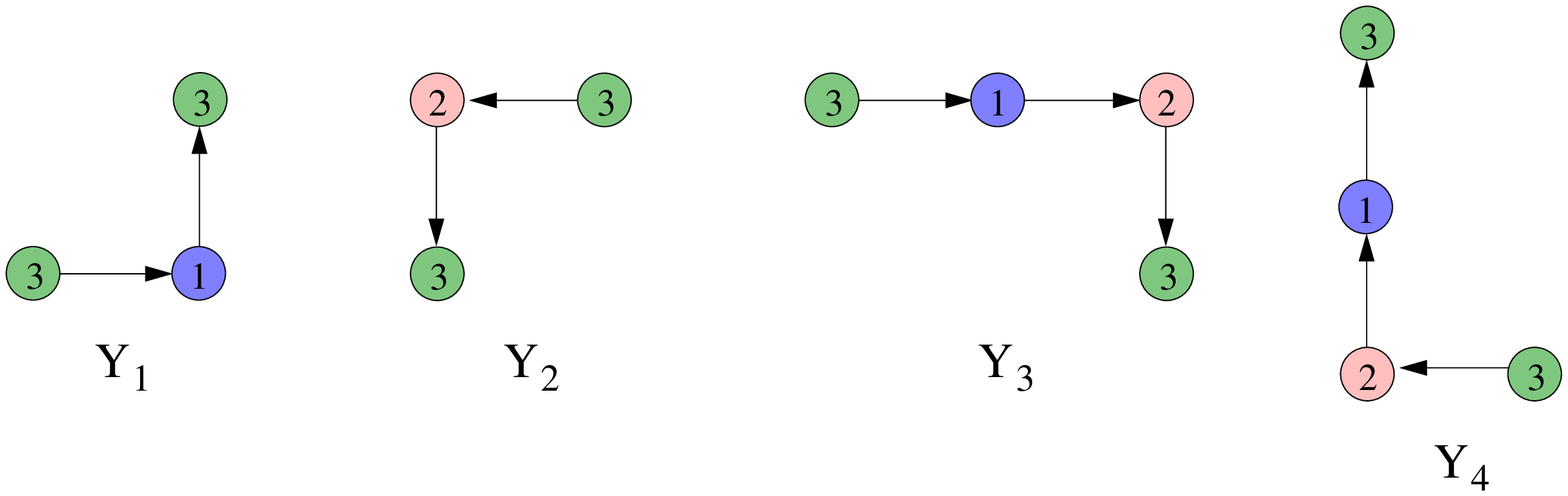}
\caption{$Y_a$ generators for the SPP.}
\label{SPP_generators}
\end{center}
\end{figure}

\begin{figure}[h]
\begin{center}
\includegraphics[width=7.5cm]{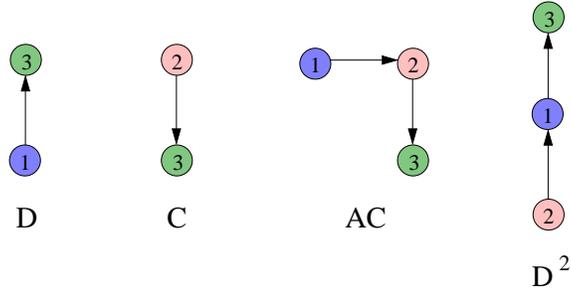}
\caption{The $s_\mu$ operators for the SPP.}
\label{SPP_tips_of_paths}
\end{center}
\end{figure}

\begin{figure}[h]
\begin{center}$
\begin{array}{ccccc}
\includegraphics[height=3.4cm]{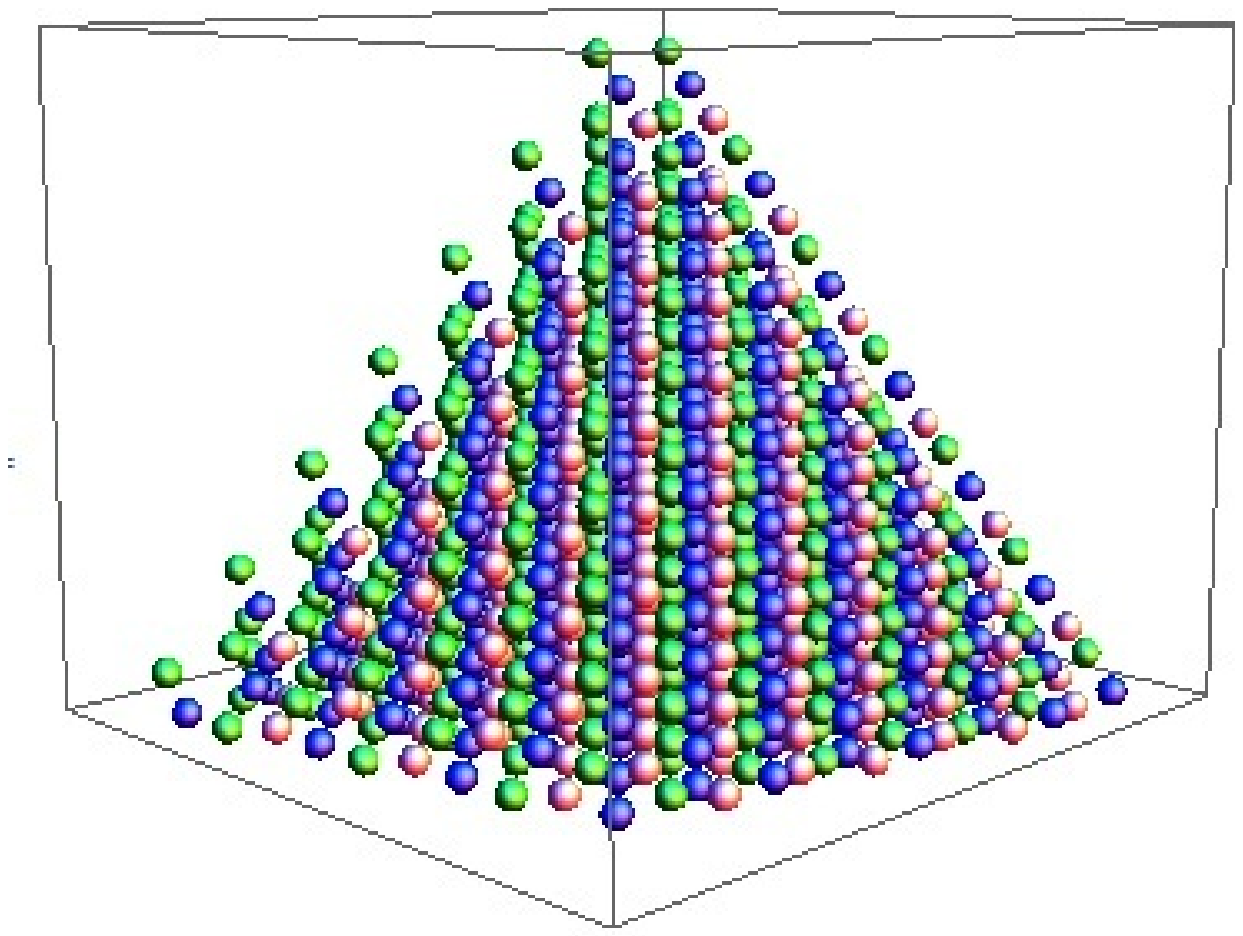} & \ &
\includegraphics[height=3.4cm]{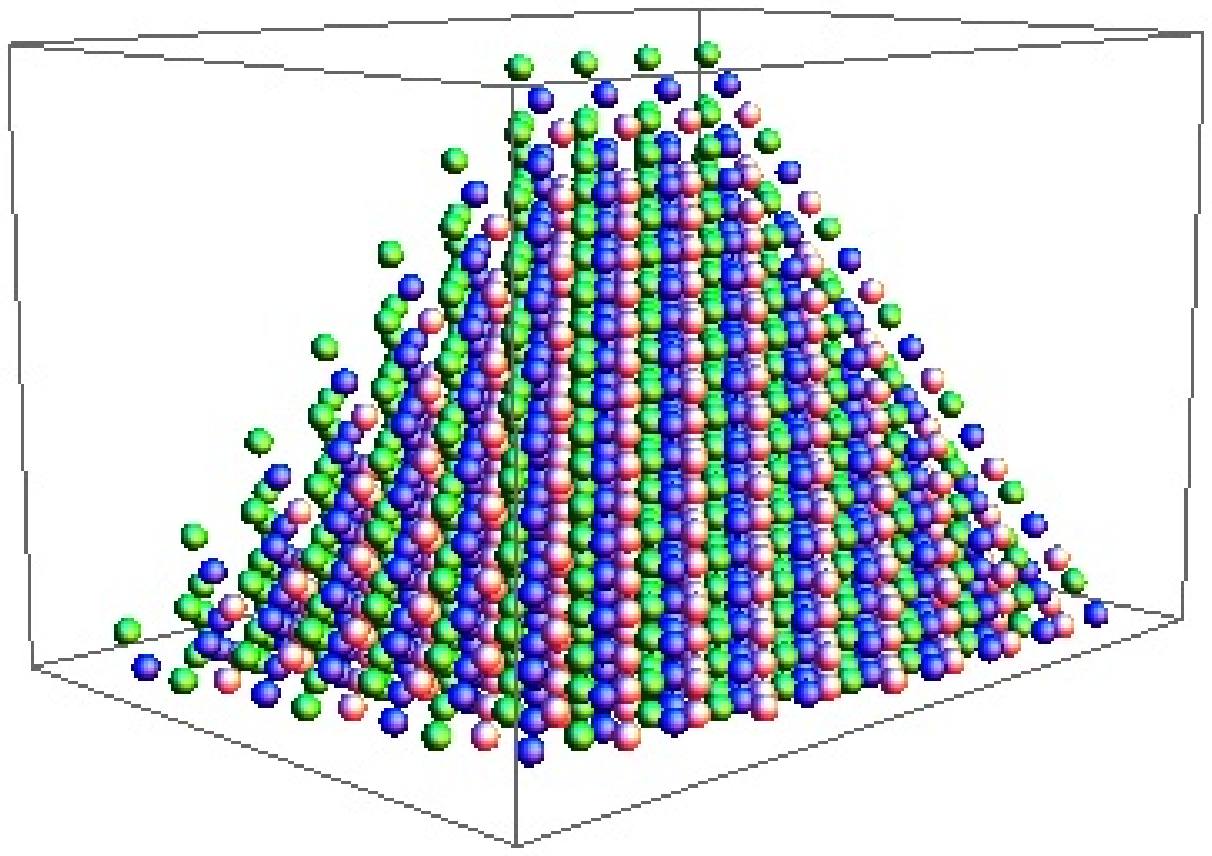} & \ &
\includegraphics[height=3.4cm]{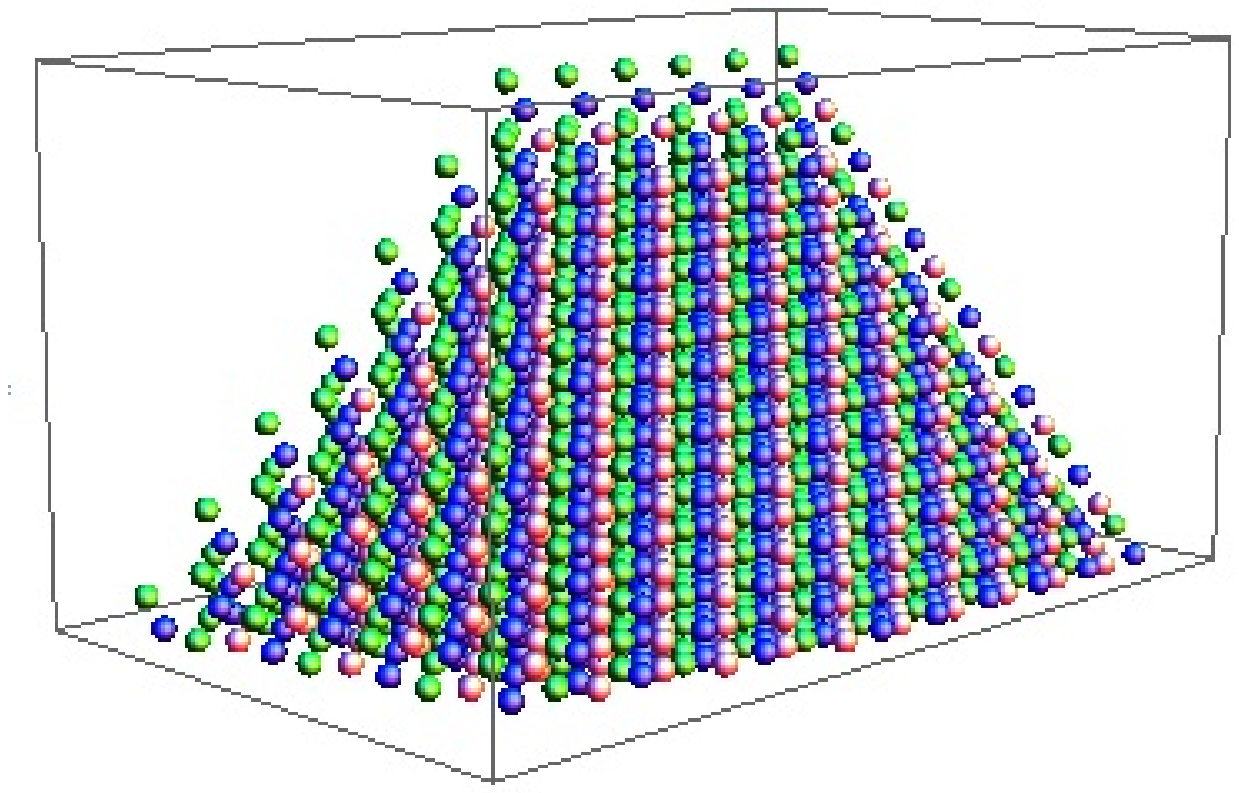} \\
n=2 & & n=4 & & n=6
\end{array}$
\end{center}
\caption{Infinite pyramids with $n=2$, $4$ and $6$ top stones for the SPP with framing superpotential corresponding to $\mathcal{O}_i=X_{32}$ and $\tilde{\mathcal{O}}_j=X_{31} X_{12}$. We display stones corresponding to paths with up to 20 bifundamental fields in the quiver.}
\label{pyramids_SPP}
\end{figure}

As we have already explained, stones in the pyramid correspond to all open oriented paths finishing with a $q_i$. An elegant way of classifying these chiral operators is via a path algebra analysis which, for the simple example at hand, can be briefly summarized in physical terms as follows: 

\begin{itemize}
\item First, identify a set of generators of paths connecting pairs of nodes of type 3, the one we use for framing. These generators correspond to the set of minimal length paths (i.e. that cannot constructed by composing shortest paths) connecting two type 3 nodes that are not equivalent under F-term relations. We denote these operators $Y_a$.

\item Next, identify similar minimal length paths that connect nodes of type 3 to all other possible nodes. We denote these operators $s_\mu$. We can reach stones of all types by inserting an $s_\mu$ at the end of a string of $Y_a$ operators.

\item Schematically, the most general operator corresponding to a stone in the pyramid takes the following form:

\beq
s_\mu^p (\prod_a Y_a) q_i
\eeq
where $p=0$ for type 3 stones and $1$ for the others.  

\end{itemize}
The $Y_a$ generators for the SPP example are shown in \fref{SPP_generators}.
\fref{SPP_tips_of_paths} shows the $s_\mu$ operators connecting node 3 to nodes 1 and 2 for the example under consideration.
We now have all necessary ingredients to construct the corresponding infinite pyramids, which are shown in \fref{pyramids_SPP} for $n=2$, $4$ and $6$ top stones.
If we project a 3d pyramid onto the horizontal plane, its edges give rise to a discretized version of the $(p,q)$ web \cite{Aharony:1997bh} dual to the toric diagram of the singularity.

\subsubsection{Finite versus infinite pyramids}

We have just seen in an explicit example how framing with $q$'s gives rise to infinite pyramids. The reason for this is that there are $n$ top stones and only $(n-1)$ relations. Inverting the roles of $p$'s and $q$'s results in $(n-1)$ top stones and $n$ relations, giving rise to finite pyramids. This can be summarized as follows:

\bigskip
\medskip

\noindent\hspace{.4cm}\begin{tabular}{|c|}
\hline
{\bf Framing with $p_j$: Finite Pyramids} \\ \\
\begin{tabular}{rl}
{\bf 1)} & Introduce flavors $q_i$ transforming in the fundamental representation of \\ & gauge groups $\alpha_i$, $(i=1,\ldots, n)$. \\ & \\

{\bf 2)} & Introduce flavors $p_j$ transforming in the antifundamental representation \\ & of gauge groups $\beta_j$, $(j=1,\ldots, n-1)$. \\ & \\

{\bf 3)} & Add to the superpotential the interactions \eref{W_framing_1_Kahler}. \\ & \\
& Every stone corresponds to an open oriented path starting with a $p_j$. Paths  \\ & containing a $q_i$ are eliminated by setting $q_i=0$, which implies that $F_{p_j}=0$ \\ & and we are left with the F-term equations of the $q_i$ as relations. There are \\ & $(n-1)$ top stones and $n$ relations, resulting in a {\it finite} pyramid.
\end{tabular} \\ 
\hline
\end{tabular}

\bigskip
\medskip

\begin{figure}[h]
\begin{center}
\includegraphics[width=5.7cm]{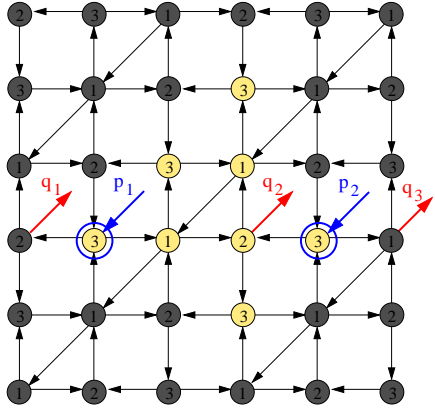}
\caption{Top view of a finite pyramid for the SPP resulting from framing with the $p_i$ flavors. Stones in the pyramid are indicated in yellow, and top stones are marked with blue circles.}
\label{periodic_quiver_SPP_finite_pyramid}
\end{center}
\end{figure}

Let us illustrate these ideas with an explicit example. \fref{periodic_quiver_SPP_finite_pyramid} shows the periodic quiver for the SPP with a flavor configuration that is obtained by starting with a single flavor and Seiberg dualizing three times.\footnote{This is an interesting example because it contains flavors for the three types of nodes. It is analogous to a flavor configuration that will be discussed later and appears in \fref{quiver_SPP_Seiberg_duality}.b.} The flavor superpotential is given by \eref{W_framing_1_Kahler}, with $\mathcal{O}_1=\mathcal{O}_2=X_{32}$, $\tilde{\mathcal{O}}_1=X_{31}X_{12}$ and 
$\tilde{\mathcal{O}}_2=X_{31}$. Framing with the $p_j$'s means that we impose the F-term equations of the $q_i$'s, which are given by

\beq
\begin{array}{l}
p_1 X_{32}=0 \\
p_1 X_{31} X_{12} = p_2 X_{32} \\
p_2 X_{31}=0
\end{array}
\eeq
As a result, we obtain a finite pyramid, whose stones are indicated in yellow in \fref{periodic_quiver_SPP_finite_pyramid}.\footnote{In general, there can be multiple stones at a given point in the periodic quiver plane, corresponding to different vertical positions. This is not the case for this example and \fref{periodic_quiver_SPP_finite_pyramid} indeed shows all the stones in the pyramid. In Section \ref{section_shadow} we will refer to this projection as the {\it shadow} of the pyramid and will introduce a method for its determination.}

\section{Pyramids and Seiberg Duality}

\label{section_pyramids_Seiberg_duality}

We have explained how to define pyramids using framing flavors and their superpotential couplings. One of the central topics of this paper is the behavior of pyramids under Seiberg duality \cite{Seiberg:1994pq}. Since we are interested in constructing pyramids, we foucs on dualizations giving rise to quivers which can be described by brane tilings. As a result, we restrict to dualizing quiver nodes that have two incoming and two outgoing arrows. In the brane tiling this type of node corresponds to a square face and Seiberg duality acts by an urban-renewal transformation \cite{Franco:2005rj}.

Let us call the dualized node $i_0$ and denote the corresponding arrows $X_{i_0,j_1}$, $X_{i_0,j_2}$, $X_{j_3,i_0}$ and $X_{j_4,i_0}$.  In addition, let us consider the case in which the node has an antifundamental flavor $p_{i_0}.$  The discussion applies to the case with a fundamental flavor $q_{i_0}$ with obvious changes. Let us consider how framing flavors transform under Seiberg duality. First, $p_{i_0}$ is replaced by a fundamental flavor of $i_0$. In addition, the following new flavors are generated as Seiberg mesons

\begin{eqnarray}
p_{j_1} & = & p_{i_0} X_{i_0,j_1}  \nonumber \\
p_{j_2} & = & p_{i_0} X_{i_0,j_2}  
\end{eqnarray}
with superpotential couplings that follow from the standard rules of Seiberg duality. If some of the new flavors become massive due to a coupling to a pre-existing one, we integrate them out using their equations of motion. In summary, Seiberg duality modifies the structure of framing flavors giving rise to a new pyramid.

Sequences of Seiberg dualities are also known as {\it duality cascades}.
Of particular interest are those cascades that are periodic. The conifold gauge theory has a well-known periodic sequence of Seiberg dualities \cite{Klebanov:2000hb}. The effect of Seiberg duality on the conifold quiver with framing flavors and its associated pyramids was investigated in \cite{Chuang:2008aw}. In this theory, Seiberg duality changes the number of top stones in the pyramids. 

\begin{figure}[h]
\begin{center}
\includegraphics[width=15cm]{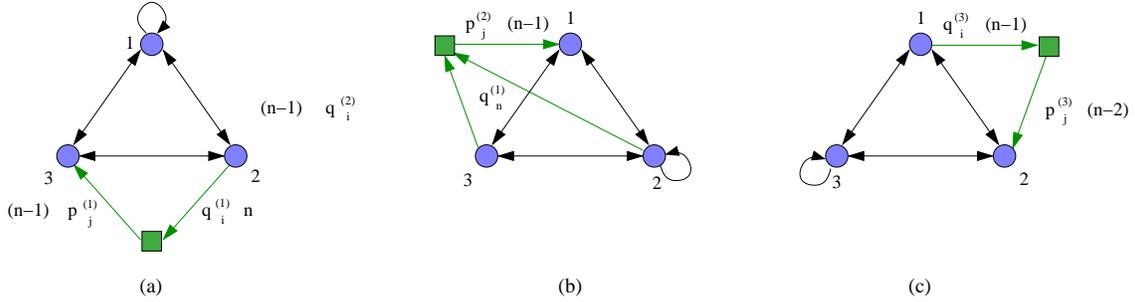}
\caption{a) The original SPP theory with framing flavors. b) The quiver after dualizing node 3. c) The quiver after further dualizing node 1. We have included a superindex to indicate the step in the dualization sequence at which flavors are generated. The final theory is identical to the original one (including superpotential coupling) after a rotation and a reduction $n\to n-1$.}
\label{quiver_SPP_Seiberg_duality}
\end{center}
\end{figure}

We will later investigate in general how other theories behave under either periodic or general sequences of Seiberg dualities. Before closing this section, let us analyze the SPP example in some more detail. A periodic sequence of Seiberg dualities for the unflavored SPP was found in \cite{Franco:2005fd} and a detailed analysis of its dynamics was given in \cite{Simic:2010ra}. It is straightforward to check that, after consecutively dualizing nodes 3 and 1, the resulting theory is the one shown in \fref{quiver_SPP_Seiberg_duality}.c. The superpotential is the one for the SPP with additional terms

\beq
W_{rels}=p_1^{(3)} \, \mathcal{O} \, q_1^{(3)}+ (p_2^{(3)} \, \mathcal{O} + p_1^{(3)} \, \tilde{\mathcal{O}}) q_2^{(3)} +\ldots+(p_{n-2}^{(3)} \, \mathcal{O} + p_{n-3}^{(3)} \, \tilde{\mathcal{O}}) q_{n-2}^{(3)}+ p_{n-2}^{(3)} \, \tilde{\mathcal{O}} \, q_{n-1}^{(3)}
\label{W_SPP_Seiberg_duality}
\eeq
with $\mathcal{O}=X_{21}$ and $\tilde{\mathcal{O}}=X_{23} X_{31}$. This means that, after a trivial rotation of the quiver, the theory is identical to the original one but with $n\to n-1$. We conclude that in this case, as it also happened for the conifold, the effect of Seiberg duality is to modify the length of the top of the pyramid.

\section{Pyramid Partitions}

\label{section_pyramid_partitions}

All the stones in a pyramid that correspond to paths passing through a given top stone $j$ define a poset $\Delta_{j}$ \cite{Mozgovoy:2008fd}. In our convention, paths terminate at top stones for infinite pyramids and start at top stones for finite ones. The full pyramid is the non-disjoint union of the contributions from all top stones $\Delta=\bigcup_{j=1}^{n}\Delta_{j}$. The {\it pyramid partitions} are in one-to-one correspondence with the ideals of $\Delta$. Let us introduce one variable $y_i$ for each gauge group in the quiver ($i=1, \ldots, N_G$, with $N_G$ the number of gauge groups in the quiver), i.e. for each type or color of stone. To every ideal $\Omega\subseteq\Delta$ we assign the weight $\prod_{i \in Q_{0}}y_{i}^{n_i}$, where $n_i$ is the number of stones of type $i$ in $\Omega$.  The {\it colored partition function} associated to a pyramid is then defined as
\beq
Z=\sum_{\Omega\subseteq\Delta}\prod_{i}y_{i}^{n_{i}} \, .
\eeq
In the partition function, linear terms correspond to the top stones and the highest order term corresponds to all the stones in the pyramid.

A practical way of keeping track of the relations in the poset $\Delta$ is by means of a Hasse diagram. For finite pyramids, an arrow in this diagram from stone $a$ to stone $b$ indicates that $a$ is on top of $b$. In fact, in this context, the arrows correspond to chiral fields in the quiver. Top stones are represented by stones without incoming arrows. For infinite pyramids, the orientation of all arrows is reversed. The rule for constructing pyramid partitions is that whenever a stone is removed from the pyramid all stones above it, i.e. all stones contained in downward paths terminating in it, should also be removed.  

\begin{figure}[h]
\begin{center}
\includegraphics[width=3.5cm]{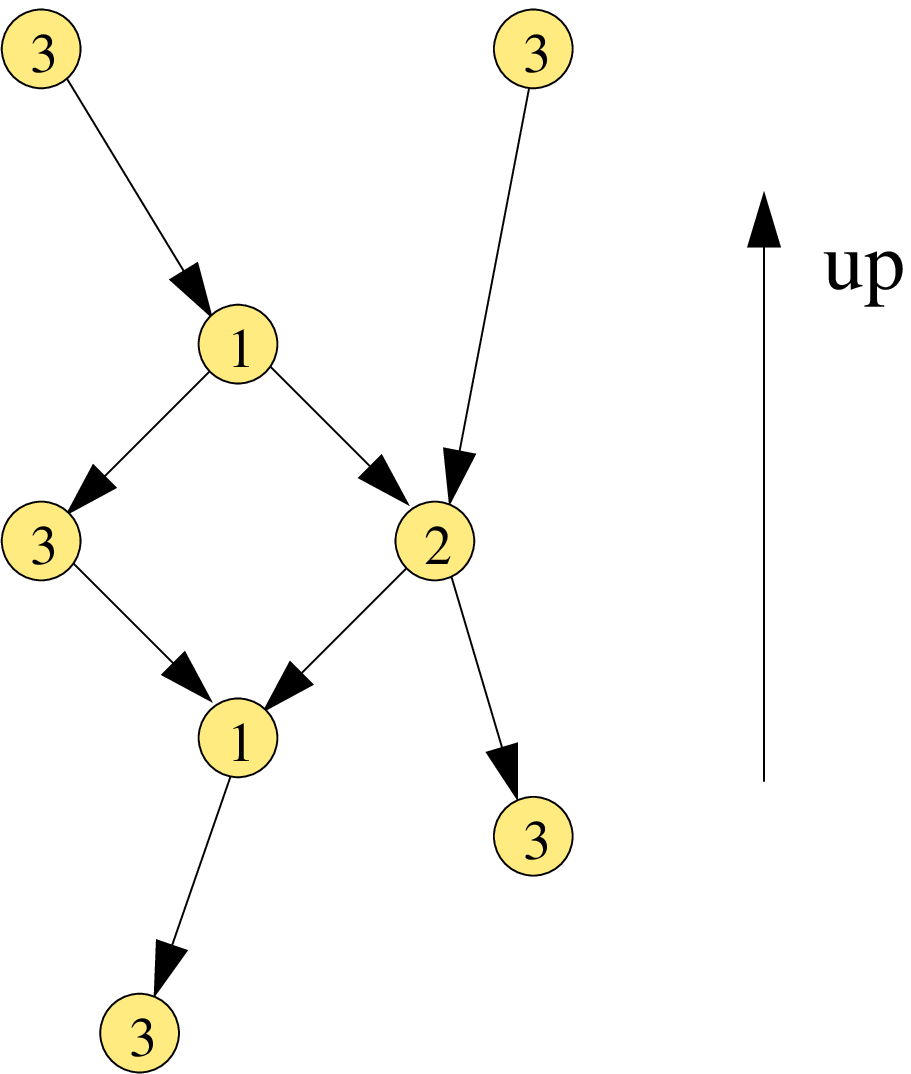}
\caption{Hasse diagram encoding the relations between stones in the poset for the pyramid in \fref{periodic_quiver_SPP_finite_pyramid}.}
\label{Hasse_SPP_finite_pyramid}
\end{center}
\end{figure}

Let us return to the example in \fref{periodic_quiver_SPP_finite_pyramid}, for which the Hasse diagram is shown in \fref{Hasse_SPP_finite_pyramid}. The partition function for this example is
\beq
Z = 1+ \underbrace{2 y_3}_\text{top stones}+y_1 y_3+ 2 y_1 y_3^2+y_1 y_3^3+y_1 y_2 y_3^2+y_1 y_2 y_3^3+y_1^2 y_2 y_3^3+2y_1^2 y_2 y_3^4+\underbrace{y_1^2 y_2 y_3^5}_\text{all stones} \, .
\eeq

\bigskip

\subsection{Connection to Generating Functions for BPS Invariants}

BPS states arising from D-branes wrapping cycles in toric Calabi-Yau manifolds 
are in one-to-one correspondence with pyramid partitions, also called crystal melting configurations \cite{Okounkov:2003sp}. The literature primarily considers toric Calabi-Yau threefolds without compact 4-cycles. These geometries give rise to quivers that, without considering framing flavors, are non-chiral. The simplest and best studied example is the conifold. Its chamber structure and the correspondence between BPS states and pyramid partitions are fully understood. The conifold chambers were originally found in \cite{pre05957514}.

\begin{figure}[h]
\begin{center}
\includegraphics[width=10cm]{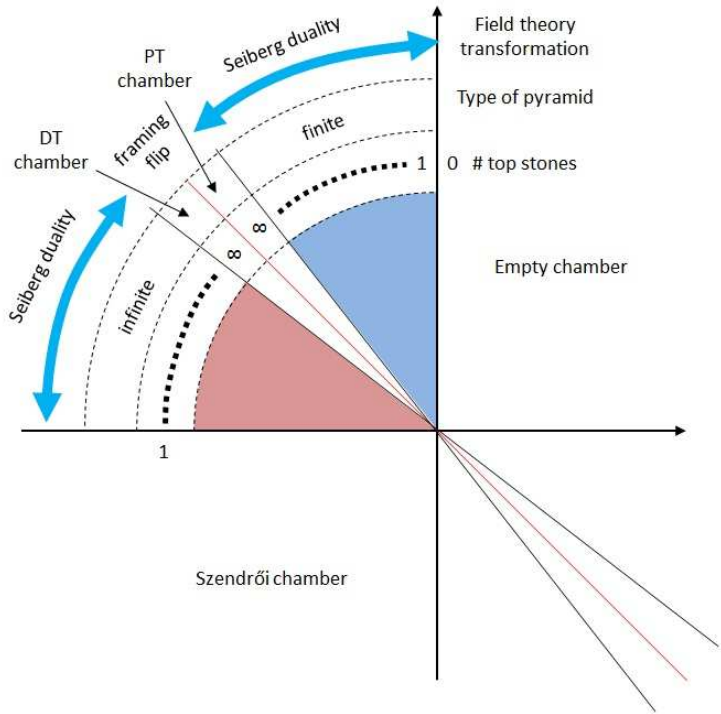}
\caption{The chamber structure for the conifold. We indicate whether the corresponding pyramids are finite or infinite and the number of stones at the top. We also indicate the gauge theory operations, Seiberg duality or framing flip, move us around chambers.}
\label{conifold_chambers}
\end{center}
\end{figure}

The connection between BPS states in the conifold and pyramids was fully investigated in \cite{Chuang:2008aw}, following previous work \cite{Szendroi:2007nu}. This paper introduced the concept of finite pyramids and studied both infinite and finite pyramids with an arbitrary number of top stones. In addition, \cite{Chuang:2008aw} also explained how all of these pyramids are constructed using quivers with framing flavors and how they are connected by Seiberg duality. These ideas combine into a compelling unified picture for the conifold that is summarized in \fref{conifold_chambers}, which shows the chamber structure, the associated pyramids, and the Seiberg duality transformations acting on flavored quivers connecting different chambers for the conifold.  

Let us start from the first quadrant in \fref{conifold_chambers}. It corresponds to the empty chamber, i.e. there are no stones in the pyramid.  The framing flavor configuration is given by one $q$, no $p$, and framing with $p$. Seiberg duality on the flavored node takes us to the second quadrant. The blue region contains an infinite number of chambers corresponding to framing with $p_i$'s, i.e. to finite pyramids. The number of top stones increases as we alternatively dualize the two nodes of the conifold quiver. An infinite number of dualities are necessary to reach the red line at 135$^{\circ}$. Immediately before this line we have the Pandharipande-Thomas chamber \cite{arXiv:0707.2348}, which has an infinite number of top stones. In order to cross the red line we flip the type of framing, going from framing with $p_i$'s to framing with $q_j$'s.  Immediately after the red line, we find the commutative Donaldson-Thomas chamber for large K\"ahler class. The red region contains an infinite number of chambers that correspond to infinite pyramids arising from framing with $q_j$'s. Alternating Seiberg dualities move us along the red region, progressively reducing the number of top stones, until reaching the third quadrant. This quadrant was studied by Szendr{\H o}i in \cite{Szendroi:2007nu} and corresponds to Donaldson-Thomas invariants in a non-commutative resolution of the conifold. The associated pyramid is infinite and has a single top stone, namely the framing is given by one $q$, no $p$, and framing with $q$. The structure in the fourth quadrant is a reflection of the one in the second quadrant.

In the previous section, we have generalized the construction of pyramids in terms of quivers with framing flavors from the conifold to arbitrary toric, singular Calabi-Yau threefolds. This approach applies to generic toric singularities, including those with compact 4-cycles, which give rise to chiral quivers, overcoming one of the main restrictions of previous analyses. Assuming that the conifold story can be extrapolated to the general geometries, it is natural to expect that the corresponding chamber structure will be a higher dimensional generalization of \fref{conifold_chambers}, with individual chambers associated to pyramids of finite or infinite type. Furthermore, we expect that lower dimensional slices of this space are described by a structure analogous to the one in \fref{conifold_chambers}. Finally, as for the conifold, the transition between different chambers would correspond to Seiberg duality on the corresponding flavored quivers. Trajectories along the mutli-dimensional space of chambers would correspond to cascades of dualities, which are the subject of Section \ref{section_duality_cascades}. We leave a more thorough investigation of the stability conditions in the general case for future work. In the next section we discuss how the partition functions for pyramids resulting from applying sequences of Seiberg dualities to quivers with framing flavors can be efficiently computed recursively using cluster transformations.

\section{Recursive Calculation of Pyramid Partition Functions}

\label{section_recursive}

\subsection{Cluster Algebras}

Cluster algebras have found applications in diverse areas of mathematics and physics since their introduction by Fomin and Zelevinsky \cite{MR1887642}.
A cluster algebra $\mathcal{A}$ of rank $n$ is a subalgebra of an ambient field $\mathcal{F} \cong \mathbb{Q}(Z_1, Z_2, \dots, Z_n)$ of rational functions in $n$ variables.  Cluster algebras are defined over a coefficient semifield $(\P, \oplus, \cdot)$. We will first define the semifield, clusters, and seeds before finally giving the definition of a cluster algebra.

We will only consider cluster algebras defined over a tropical semifield.
The tropical semifield is defined by the free group generated by variables $u_j$ indexed by a finite set $J = \{1,2, \dots, n \}$.  The multiplication operation is the usual multiplication of polynomials, while the tropical addition operation $\oplus$ is defined by
$$\prod_{j} u_j^{a_j} \oplus \prod_{j} u_j^{b_j} = \prod_{j} u_j^{\min(a_j,b_j)}$$
Cluster algebras have a distinguished set of generators called {\it cluster variables}.

A {\it labeled seed} is a triple $(\mathbf{Z},\mathbf{x},B)$ consisting of \footnote{Our notation for the cluster variables $(\mathbf{Z},\mathbf{x},B)$ is slightly different from the standard one $(\mathbf{x},\mathbf{y},B)$. We hope the reader is not confused by this choice.}

\medskip
\begin{itemize}
\item $\mathbf{Z} = (Z_1, \dots, Z_n)$ a cluster,
\item $\mathbf{x} = (x_1, \dots x_n)$ an $n$-tuple of coefficients,
\item $B = (b_{ij})$ an $n \times n$ integer matrix that is skew-symmetrizable.
\end{itemize}
\medskip

A matrix $B$ is skew-symmetrizable if there exist positive integers $d_1, \dots , d_n$ such that $d_i b_{ij} = -d_j b_{ji}$, where there is no sum over repeated indices. The matrix $B$ naturally gives rise to a quiver through its positive entries: for any two vertices $i \neq j$, there are $[b_{ij} ]_+$ arrows from $i$ to $j$ in the quiver.  
Given a labeled seed, we define the mutation of the seed in direction $k$ as follows,

Under quiver mutation, a cluster seed transforms as follows. For the matrix $B$, we have

\beq
b_{ij}' = 
\begin{cases}
-b_{ij} & \text{if } i = k\text{ or } j=k \\
b_{ij}+\text{sgn}(b_{ik}) [b_{ik}b_{kj}] & \text{otherwise}
\end{cases}
\label{b_transformation}
\eeq
where there is no sum over $k$. The coefficients transform as
\beq
x_j' = 
\begin{cases}
x_k^{-1} & \text{if } j = k, \\
x_j \prod_{arr(k \rightarrow j)} x_k  & \text{if } j \neq k \text{ and } x_k \text{ has positive exponent,} \\
x_j \prod_{arr(j \rightarrow k)} x_k  & \text{if } j \neq k \text{ and } x_k \text{ has negative exponent.}
\end{cases}
\label{y_transformation}
\eeq
Finally the cluster variables transform according to
\beq
Z_k' = \frac{\prod_{arr(k \rightarrow j)}Z_j+x_k \prod_{arr(j \rightarrow k)} Z_j}{(x_k \oplus 1) Z_k}.
\label{Z_transformation}
\eeq
The tropical sum in the denominator takes care of the two possibilities in \eref{y_transformation}, namely whether $x_{k}$ has positive or negative exponents. Notice that \eref{y_transformation} and \eref{Z_transformation} take into account all arrows coming in or out of node $k$. In particular, these transformation rules also incorporate the effect of possible pairs of bi-directional arrows, which have a zero net contribution to the matrix $B$. The matrix $B$ is not sufficient for keeping track of all arrows and we have to follow the rules of Seiberg duality instead. In particular, the quiver might contain bi-directional arrows, which do not contribute to $B$.

Finally we can define a cluster algebra as follows. Starting from an initial seed, consider all possible mutations. In the absence of bi-directional arrows, this the union of all of the clusters obtained from these mutations defines a {\it cluster algebra}. The general quivers with superpotentials we study in this paper give rise to structures that are slightly more general. As already implicit in the previous paragraph the superpotential is crucial for keeping track of the number of oriented arrows in the quiver adjacency matrix.

\subsection{Physical Interpretation in Terms of Flavored Quivers}

As we have already mentioned, this setup has a natural interpretation in terms of quiver gauge theories. In this paper, we will restrict to cases in which the matrix $B$ is antisymmetric, i.e. skew-symmetric. In this case, $B$ can be thought of as the anti-symmetrized adjacency matrix of the quiver. 
Mutations of a labeled seed correspond to a mutation of the quiver $Q$, i.e. to a Seiberg duality transformation. 

The transformation rules in \eref{y_transformation} suggest that we can identify the coefficients $x_k$ with the exponential of the fractional brane charges for each of the gauge groups. As we perform mutations, \eref{y_transformation} indicates how to express the new fractional brane charges in terms of those of the original quiver. We refer the reader to \cite{Feng:2002kk} to a discussion of mutations in this context.

One of the main points we emphasize in this paper is that the variables $\mathbf{Z}$ can be identified with pyramid partition functions. 

\begin{center}
\begin{tabular}{|c|}
\hline 
 \ \ Cluster Variables $\mathbf{Z}$ = Pyramid Partition Functions \ \ \\
\hline
\end{tabular}
\end{center}
\noindent This identification applies to both finite and infinite pyramids and allows a very efficient recursive computation of the corresponding partition functions using \eref{y_transformation} and \eref{Z_transformation}.

In all the examples we will consider, the $x_j$ coefficients will only contain positive exponents of the initial gauge group variables $y_i$. In this case, the transformation rules become
\beq
x_j' = 
\begin{cases}
x_k^{-1} & \text{if } j = k, \\
x_j \prod_{arr(k \rightarrow j)} x_k  & \text{if } j \neq k ,
\end{cases}
\eeq
and
\beq
Z_k' = \frac{\prod_{arr(k \rightarrow j)}Z_j+x_k \prod_{arr(j \rightarrow k)} Z_j}{Z_k}.
\eeq
It is important to keep in mind that \eref{y_transformation} and  \eref{Z_transformation} provide the transformation rules for the general case.

There is no reference to framing flavors in \eref{y_transformation} and \eref{Z_transformation}, which might lead us to incorrectly think that framing flavors play no role in this formalism. There are different ways in which these equations can be used to generate partition functions recursively. In the case of finite pyramids, it is natural to start from trivial initial data, where the partition functions for all nodes correspond to empty room configurations, i.e. $Z_k=1$ for all $k$. This assumption indeed constrains the configuration of framing flavors, since we have to guarantee that the flavors are such that $Z_k=1$ if we dualize the $n$ gauge groups in inverse order. In this case, the pyramid partition functions are equal to the $F$-polynomials defined in \cite{MR2295199}.
For both finite and infinite pyramids, a more general option is to apply \eref{y_transformation} and \eref{Z_transformation} to arbitrary flavor configurations. As in the previous case, we need to specify the initial conditions of the recurrence, given by the $Z_k$'s for all $k$. These partition functions can be obtained by direct computation using the flavor configurations that result from the desired one after dualizing all nodes in the quiver.

\subsection{Finding the Shadow of a Pyramid}

\label{section_shadow}

A useful concept when dealing with finite pyramids is that of the {\it shadow} of the pyramid. The shadow is the vertical projection of the pyramid onto the plane of the periodic quiver. 

Here we introduce a recursive procedure for directly constructing the shadow, without need for constructing the full pyramid. The starting point is a quiver and superpotential $(Q,W).$ Let the initial partition function for each node be 1 and all coefficients be $x_\mu=y_\mu$, where $\mu$ indicates the corresponding node of the quiver. 
We want to determine the shadow of the pyramid associated to the quiver that results from applying a sequence of toric mutations, i.e. Seiberg dualities, to this configuration.

Now pick a node in the quiver obtained after all the dualities and add a single outgoing framing arrow. We will explain how the partition function associated to this node gives rise to the shadow of the resulting pyramid. We initially start with no marked nodes.

\begin{itemize}
\item Each time we dualize a node $\mu$ with an outgoing framing arrow we replace it by its Seiberg dual $\hat{\mu}$ and add $\hat{\mu}$ to the collection of marked nodes.  We then add an incoming framing arrow to $\hat{\mu}.$
We next consider nodes  $\mu'$ with an arrow pointing towards $\mu.$
If $\mu'$ does not have an incoming framing arrow we add an outgoing framing arrow to $\mu'$.
If $\mu'$ has an incoming framing arrow, we simply delete the incoming framing arrow. 

\item For each node $\mu$ we dualize that has an arrow pointing to a marked node, we add its dual $\hat{\mu}$ to the set of marked nodes.
\end{itemize}

\begin{figure}[h]
\begin{center}
\includegraphics[width=7.5cm]{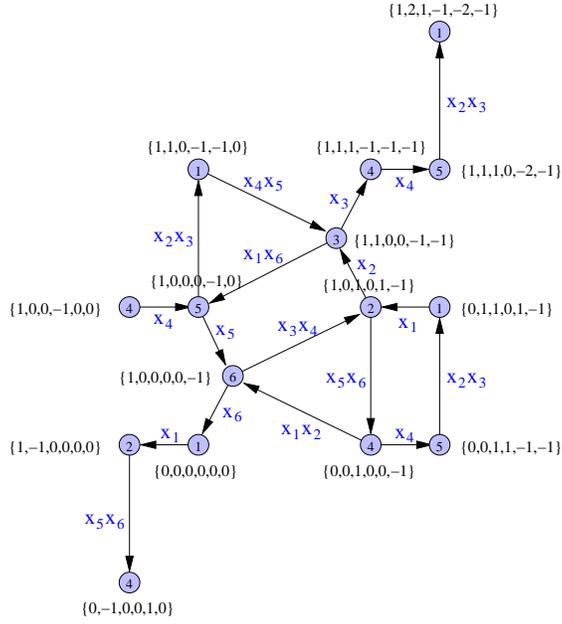}
\caption{The shadow of a pyramid for $dP_3$. We indicate the $\Psi$-coordinates for the quiver nodes and the homomorphisms associated to the arrows. This figure is the dual graph to the diamond of order 2 in \cite{Cottrell:2010my}.}
\label{fig:dP3shadow}
\end{center}
\end{figure}

Let us illustrate this construction with an explicit example for $dP_3$. We start from its phase I, which is given in Appendix \ref{section_dP2_and_dP3}. Next, we Seiberg dualize the gauge groups $(1,4,2,5,3,6,4)$, resulting in a new periodic quiver. We then place a single outgoing framing arrow on one of the nodes of the final periodic quiver. In this example, we place an outgoing framing arrow on node $4$, whose coordinates are $\{2,0,-1,1,0-2\}.$  Since this node has an outgoing framing arrow, it is not marked. We then apply the same sequence of Seiberg dualities, but in reversed order, i.e. $(4,6,3,5,2,4,1)$, and update the framing arrows and marked nodes according to the procedure described above. The shadow of the pyramid is the final collection of marked nodes on the original periodic quiver. The result is shown in \fref{fig:dP3shadow} and corresponds to one of the steps in the general sequence of $dP_3$ duals that we study in Section \ref{section_recursive_dP3}. We provide the explicit expression for its partition function in Appendix \ref{section_partition_functions}. It corresponds to $Z_7$ in the $dP_3$ sub-section. 

Physically, the construction of the shadow, and in fact the determination of the entire pyramid, should be independent of the sequence of Seiberg dualities used to arrive at the final theory. If two different sequences of Seiberg dualities lead to a node with the same coordinates, a conjecture about cluster algebras we can put forward is that the partition functions obtained for this node by the two different sequences agree.

\subsection{Connection with the Multidimensional Octahedron Recurrence}

We will now show that the recurrence equation for pyramid partition functions can be uniformly described as the {\it multidimensional octahedron recurrence} \cite{MR2317336,MR2261754,MR2565555}. In short, if we label quiver gauge groups using $\Psi$-coordinates subject to a constraint that we discuss below, the cluster transformation \eref{Z_transformation} becomes the multidimensional octahedron recurrence.

The multidimensional octahedron recurrence for any toric Calabi-Yau singularity with $D$ points on the boundary of its toric diagram can be embedded into a $(D - 1)$-dimensional lattice. The partition functions $Z_I$ are labeled by an index $I \in \mathbb{Z}^D$ subject to the constraint that $\sum_{i=1}^D I_i = 0.$ In section \ref{section_R_charge}, we showed that the $\Psi$-map provides a unique coordinate $I_{\mu} \in \mathbb{Z}^D$ for each node $\mu$ of the periodic quiver up to an integral multiple of $\sum_{i=1}^D E_i.$ Hence, it is natural to project from the $\mathbb{Z}^D$ lattice given by the $\Psi$-coordinates down to a sub-lattice $\Lambda \subset \mathbb{Z}^D$, consisting of vectors $c_i \in \mathbb{Z}^D$ such that $\sum_{i = 1}^{D} c_i = 0.$ More concretely, in terms of coordinates, we can choose the map $\mathbb{Z}^D \rightarrow \Lambda$ to take
the vectors $(b_1, b_2, \dots, b_D)\in \mathbb{Z}^D$ to the vectors $(b_1 - b_2, b_2 -b_3, \dots, b_{D} - b_1) \in \Lambda.$

Under the $\Psi$-map, every arrow in the quiver is a sum over sets of consecutive points on the boundary of the toric diagram. After projecting the corresponding vector down to sub-lattice $\Lambda$, every arrow takes the form $\pm(e_i - e_j).$ 
The cluster transformation for partition functions \eref{Z_transformation} takes the form
\beq
Z_k' Z_k = \frac{\prod_{arr(k \rightarrow j)}Z_j+x_k \prod_{arr(j \rightarrow k)} Z_j}{(x_k \oplus 1)} \, .
\eeq
We are interested in dualizing toric nodes, which have for arrows that alternate between incoming and outgoing. We label them as $e_i - e_j$, $-e_j + e_k$, $e_k - e_l$ and $-e_l + e_i$, where $i$, $j$,$k$, and $l$ are in cyclic order. Rewriting the cluster transformation for toric nodes in terms of the lattice $\Lambda$, we obtain
\beq
Z_{I + e_i + e_k - e_j - e_l} Z_{I} = 
\frac{Z_{I + e_i - e_l} Z_{I + e_k - e_j} + x_k Z_{I + e_k - e_l} Z_{I + e_i - e_j}}{(x_k \oplus 1)} \, .
\eeq
Thus, we have succeeded in writing the cluster transformation for toric nodes in terms of the multidimensional octahedron recurrence.

\bigskip

\section{Duality Cascades}

\label{section_duality_cascades}

We are interested in the behavior of quiver theories and their associated pyramids under Seiberg dualities.  Additionally, we conjecture that Seiberg dualities generate transitions between different stability chambers. In this section we discuss various perspectives on duality cascades which are sequences of Seiberg dualities. 

\subsection{Geometry and Duality Cascades}

\label{section_duality_geometry}

We now review the connection between the Calabi-Yau geometry, the space of Seiberg dual theories, and the cascades of dualities that generate translations in this space. 
Consider a toric Calabi-Yau threefold with a toric diagram whose perimeter is equal to $D$. The global symmetry of the associated quiver theory contains a $U(1)^3$ subgroup coming from the isometries of the toric Calabi-Yau. One linear combination of these $U(1)$'s is the superfoncormal R-symmetry. The global symmetry group can contain additional $U(1)$ factors, whose corresponding bulk gauge fields come from the reduction of the Ramond-Ramond 4-form potential $C_4$ over 3-cycles. These $U(1)$ symmetries are called baryonic because D3-branes are charged under $C_4$ and D3-branes wrapped over supersymmetric 3-cycles give rise to dibaryonic states in the gauge theory. The number of independent 3-cycles in $\mathcal{C}$ is $(D-3)$. The global symmetry group hence contains the following subgroup
\beq
U(1)_R\times U(1)_F^2 \times U(1)_B^{D-3} \, .
\eeq

Fractional branes correspond to D5-branes wrapping compact 2-cycles. They modify the ranks of gauge groups in the dual quiver, breaking conformal invariance and inducing a Renormalization Group (RG) flow that takes the form of a duality cascade. There are as many independent fractional branes as baryonic $U(1)$ symmetries. In fact, it is possible to use each $U(1)_B$ to determine a rank vector through the prescription given \cite{Benvenuti:2004wx}, which we now review. We initially set all quiver ranks equal to $N$, corresponding to the absence of fractional branes. Next, we choose a node $I$ and change its rank $N_I$ from $N$ to $N+M$, where $M$ is the number of fractional branes of a given type. Then, we pick a bifundamental arrow going from node $I$ to a node $J$. The rank of node $J$ is $N_J=N_I+U(1)^{I\to J}_B M$, where $U(1)^{I\to J}_B$ is the integer baryonic charge of the $I\to J$ bifundamental. The process is repeated until determining rank assignments for all gauge groups. This physical procedure for determining rank assignments associated to fractional branes is indeed equivalent to the more formal discussion in Section \ref{section_quivers_and_geometry}. The modules $T(\hat{b}_i)$ give the ranks for gauge group $i$ for all $(D-3)$ possible fractional branes. As a concrete example, the modules in \fref{fig:dP2quiver} can be re-interpreted as fractional branes. 

While we are not interested in RG flows in this paper, i.e. we are not going to restrict ourselves to sequence of dualities driven by the beta functions for the gauge couplings, the previous discussion makes it clear that the ``dimension" of the space of cascades is equal to $(D-3)$.  
In addition, the $U(1)_B^{D-3}$ symmetry identifies directions in the space of dual theories associated to simple cascades, i.e. those associated to RG flows for the corresponding fractional branes.

\subsection{Duality Cascades from Zonotopes}

\label{section_duality_zonotopes}

\begin{figure}[h]
\begin{center}
\includegraphics[width=6.5cm]{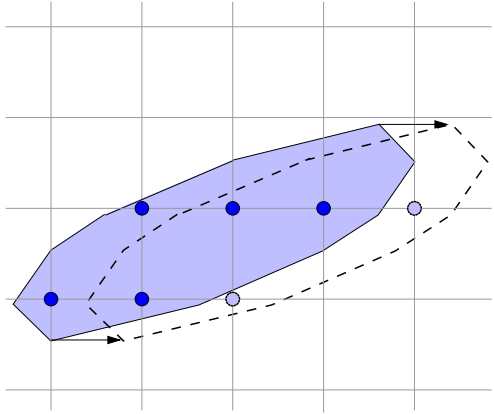}
\caption{Motion of the zonotope in the $(D-3)$ dimensional ambient space gives rise to duality cascades. We can define basic periodic cascades as translations along the different axes.}
\label{zonotope_duality}
\end{center}
\end{figure}

The zonotope construction of Section \ref{sec:zonopq} provides an efficient tool for identifying cascades. Translations of a zonotope in its $(D-3)$ dimensional ambient space generate  Seiberg dualities. As we explained in Section \ref{sec:zonopq}, internal points correspond to gauge groups in the quiver gauge theory. As a zonotope is shifted, a new lattice point enters the zonotope every time another lattice point lives it. These two points represent the same gauge group before and after Seiberg duality. \fref{zonotope_duality} shows this process for $dP_2$. This description gives rise to a natural basis for the space of cascades, analogous to the one in the previous subsection, with `basic' periodic cascades associated to the motion along each of the axis of the $(D-3)$ dimensional space.

\subsection{Duality Webs}

The connections between gauge theories related by Seiberg dualities can be nicely encoded by a {\it duality web} \cite{hep-th/0110028,hep-th/0306092}. Every node in a duality web represents a gauge theory. A link between two nodes indicates that the corresponding gauge theories are connected by a Seiberg duality. Different types of nodes correspond to gauge theories with different quivers and superpotentials. Distinct nodes of the same type correspond to gauge theories that differ only by a permutation of their gauge groups. A closed loop in a duality web indicates a sequence of dualities that comes back exactly to the same theory. Duality webs are constructed based on the un-flavored quivers. After including framing flavors, points in the web correspond to quivers with different flavor structures and thus give rise to pyramids of varying size, which are infinite or finite depending on the framing choice.

As already mentioned, in this paper we restrict ourselves to toric quivers, i.e. those that can be completely encoded by periodic quivers or, equivalently, dimer models. These are theories in which all gauge groups have rank $N$ when only $N$ D3-branes are present. This means that we only consider the dualization of gauge groups with two incoming and two outgoing arrows (i.e. square faces in the tiling). We will refer to the corresponding nodes in the quiver as {\it toric nodes}. Below, we discuss the cases of $dP_2$ and $dP_3$. Notice that this description of the space of dual theories is more refined than the one in the previous sub-section. In particular, at any point of the duality web the number of toric nodes (and hence directions in which one can possible move) is generically greater than the dimension of the space of periodic cascade, which is equal to $(D-3)$.

\bigskip

\subsection*{Del Pezzo 2}

\begin{figure}[h]
\begin{center}
\includegraphics[width=8cm]{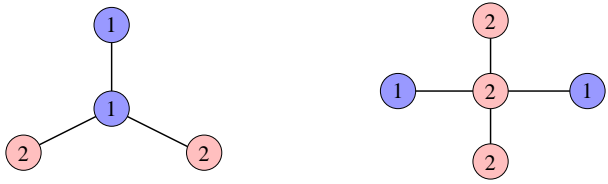}
\caption{Elementary nodes in the toric duality web for $dP_2$, indicating how phases I and II transform under Seiberg duality.}
\label{nodes_web_dP2}
\end{center}
\end{figure}

The second del Pezzo has two toric phases which are described in Appendix \ref{section_dP2_and_dP3}.  Phases I and II of $dP_2$ have three and four toric nodes respectively. \fref{nodes_web_dP2} shows how the two toric phases are transformed under all possible toric dualities. The number of lines emanating from each phase is equal to the number of toric nodes in that quiver. The duality web is constructed by gluing these two elementary building blocks.

\begin{figure}[h]
\begin{center}
\includegraphics[width=12cm]{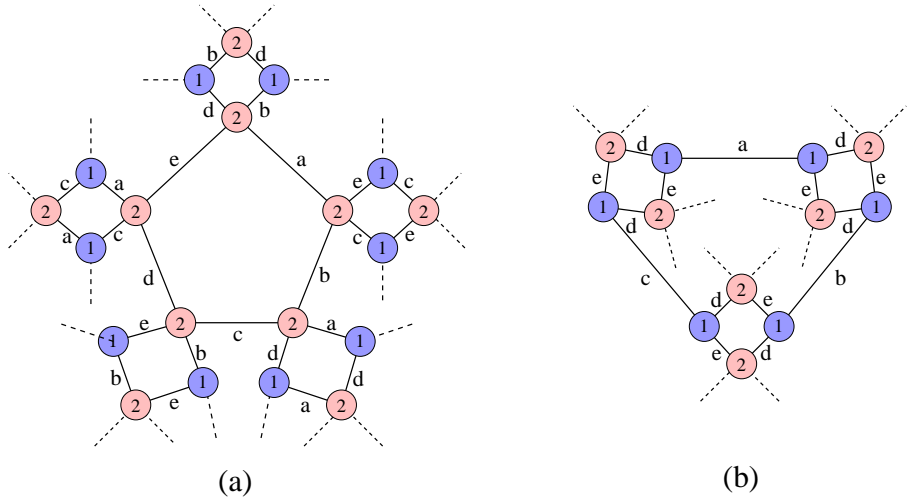}
\caption{The two basic sub-structures that form the toric duality web for $dP_2$. Letters over links indicate the dualized gauge groups and can take any of the possible values $1,\ldots,5$. Different letters correspond to different numerical values.}
\label{dP2_web_building_blocks}
\end{center}
\end{figure}

Some interesting conclusions can be already drawn from \fref{nodes_web_dP2}. We see that it is impossible to have a cascade involving only phase I, while it is possible to have one that uses only phase II. Such cascade has been studied in detail in \cite{Franco:2005fd}.
The $dP_2$ web of toric duals has a very rich structure which can be completely charted. Remarkably, it can be built by combining two types of sub-structures shown in \fref{dP2_web_building_blocks}.

\bigskip

\subsection*{Del Pezzo 3}

\begin{figure}[h]
\begin{center}
\includegraphics[width=15cm]{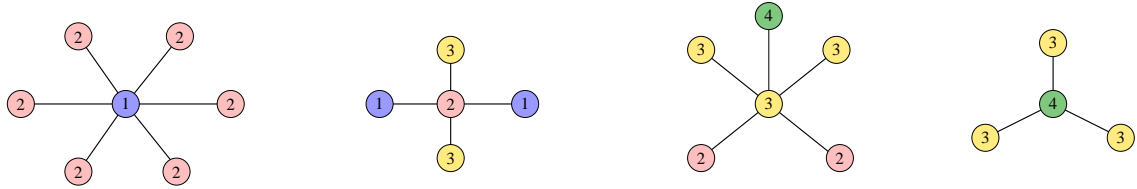}
\caption{Elementary nodes in the toric duality web for $dP_3$, indicating how its four toric phases transform under Seiberg duality.}
\label{nodes_web_dP3}
\end{center}
\end{figure}

The third del Pezzo has four toric phases, which we summarize in Appendix \ref{section_dP2_and_dP3}. Once again, it is useful to classify how each phase transforms when dualizing toric nodes. The result is shown in \fref{nodes_web_dP3}.
As in the $dP_2$ case, we can already derive useful conclusions from \fref{nodes_web_dP3}. First, we can only construct a cascade that uses one phase for phase III. Careful analysis of the resulting sequence of dualities shows that this cascade is not too interesting from the finite pyramid point of view (in particular, pyramids do not change size). We also see that there is a cascade that alternates phases I and II. This cascade is analyzed in Section \ref{section_recursive_dP3}.

\bigskip

\subsection{Duality Webs versus Zonotopes}

Zonotopes and duality webs provide complementary characterizations of the space of dual theories. Here we present a brief comparison of both approaches.
From the discussion Sections \ref{section_duality_geometry} and \ref{section_duality_zonotopes}, we conclude that the dimensionality of the space of cascades is $(D-3)$.
Zonotopes provide a graphical way of understanding this. Cascades correspond to translations of the zonotope in a $(D-3)$ dimensional space. As the zonotope is translated, only gauge groups associated to points close to the boundary of the zonotope are dualized.
On the other hand, duality webs represent the space of all possible Seiberg duals. Duality webs map the entire space of dual theories and cluster transformations allow us to determine of the partition functions for all of them. 

\bigskip

\section{Recursive Calculation of Partition Functions: Explicit Examples}

\label{section_explicit_examples}

The discussion in Section \ref{section_recursive} is completely general and it applies to arbitrary series of Seiberg dualities acting on gauge theories associated to general toric geometries. Below we demonstrate how cluster transformations can be used to determine pyramid partition functions in some concrete examples. In all cases we start from initial conditions such that $Z_n=1$ for all $n$. At the starting point, we also identify the prefactors of the octahedron recurrence with the gauge group variables, i.e. we set $x_n=y_n$ for all $n$.

\subsection{$L^{a,b,c}$ Geometries}

Let us consider the infinite family of real cones over $L^{a,b,c}$ manifolds. These geometries were introduced in \cite{hep-th/0504225,hep-th/0505027}, and the corresponding quiver theories were found in \cite{Franco:2005sm,Butti:2005sw,Benvenuti:2005cz}. This family contains the $Y^{p,q}$ theories as a subset, which follow from setting $a=p-q$, $b=p+q$ and $c=p$. The geometries correspond to a GLSM with single charge vector $(a,-c,b,-d)$ and the associated quivers were introduced in \cite{Franco:2005sm,Butti:2005sw,Benvenuti:2005cz}. For these theories, there exists a natural ordering of gauge groups that gives rise to a duality cascade in which every step involves the same quiver, up to permutation of its nodes \cite{Eager:2010ji}.\footnote{As in other examples, other sequences of dualities are possible and cluster transformations can also be used to determine the corresponding partition functions. Moreover, these geometries generically have other toric phases in addition to the one involved in the cascade we focus on. For example, a detailed analysis of the structure of toric duals for $Y^{p,q}$ can be found in \cite{Benvenuti:2004wx}.} The $Y^{p,p-1}$ and $Y^{p,1}$ cascades analyzed in detail in \cite{Herzog:2004tr} are examples of this sequence. Focusing on this cascade, the recursive equation \eref{Z_transformation} takes the form 
\beq
Z_n Z_{n-N} = \, Z_{n-a}Z_{n-N+a}+x_{n} \, Z_{n-c}Z_{n-N+c}
\label{octahedron_recurrence}
\eeq
where $N=a+b$ is the number of gauge groups in the corresponding quiver and

\beq
x_{n}=\prod_{i=1}^{a+b} y_i^{g_{n-i}}  \ \ \ \ \ g={1\over (1-q^a)(1-q^b)} \, ,
\label{Labc_prefactor}
\eeq
where $g=\sum_n g_n q^n$. In Appendix \ref{section_partition_functions} we provide explicit partition functions for the first steps in the sequence for $dP_1=L^{1,3,2}$. Setting $y_i=1$, the $dP_1$ partition functions count the number of pyramid partitions and reduce to the Somos-4 sequence:
\beq
\text{{\bf Somos-4 sequence: }}2,3,7,23,59,314,1529,8209,\ldots
\nonumber
\eeq

\noindent These numbers illustrate the rapid growth in the number of pyramid partitions. The Somos-4 sequence has already appeared in this context in \cite{Fordy:2009qz}, where it was obtained from a graph, which is precisely the $dP_1$ dimer model. Closer to our work, the Somos-4 sequence has also been seen to arise from the octahedron recurrence on the $dP_1$ quiver in \cite{MR2317336}. It is important to emphasize that this counting is independent of introducing the prefactors \eref{Labc_prefactor}, which is crucial for our full partition functions.

\bigskip

\subsection{The $L^{a,b,a}$ Sub-Family: Factorized Partition Functions from Non-Chiral Quivers}

\label{section_non_chiral_quivers}

\begin{figure}[h]
\begin{center}
\includegraphics[width=6cm]{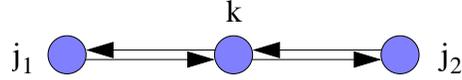}
\caption{A piece of a larger quiver showing a node to be dualized in the cascades for $L^{a,b,a}$ and $\mathbb{C}^3/(\mathbb{Z}_2 \times \mathbb{Z}_2)$ theories.}
\label{non-chiral_node}
\end{center}
\end{figure}

The set of all toric geometries without vanishing 4-cycles, i.e. those giving rise to non-chiral quivers, consists of the infinite family of real cones over $L^{a,b,a}$ geometries and $\mathbb{C}^3/(\mathbb{Z}_2 \times 
\mathbb{Z}_2)$. These geometries are also called generalized conifolds and their corresponding gauge theories and brane tilings can be found in \cite{Franco:2005sm} for example. Here we focus on the specific cascade introduced in the previous section. We immediately see that the partition function takes a product form. The reason for this is that the node $k$ that is dualized at each step has the form shown in \fref{non-chiral_node}, i.e. for each arrow $k\to j$ there is an arrow $j\to k$ connecting the same pair of nodes in the opposite direction. 
In this case the recurrence equation \eref{Z_transformation} reduces to
\beq
{Z_n Z_{n-(a+b)}\over Z_{n-a} Z_{n-b}} = (1+x_n) \, ,
\label{recurrence_Laba}
\eeq
where the coefficients $x_n$ have the closed-form expression
\beq
x_{n}=\prod_{i=1}^{a+b} y_i^{g_{n-i}}  \ \ \ \ \ g={1\over (1-q^a)(1-q^b)} \, ,
\eeq
with $g=\sum_n g_n q^n$. We can provide closed expressions for the solutions of \eref{recurrence_Laba}. They are given by
\beq
Z_n=\prod_{j=1}^n (1+x_j)^{g_{n-j}} \, .
\eeq
So we have shown that the partition functions factorization for every cascade in which the dualized node at every step takes the form in \fref{non-chiral_node}.

\subsection{Del Pezzo 3}

\label{section_recursive_dP3}

Let us now apply the cluster transformation ideas to $dP_3$. We focus on a cascade that alternates between phases I and II, which is obtained by starting from phase I as shown in \fref{quivers_dP3} and repeating the sequence of dualities $(1,4,2,5,3,6,4,1,5,2,6,3)$ on the corresponding gauge groups.
In Appendix \ref{section_partition_functions} we present the full partition functions for the first steps in the cascade. Odd steps correspond to phase I and even ones correspond to phase II. Consecutive phase II and I partition functions are equal, i.e. $Z_{2p}=Z_{2p-1}$, after the change of variables  $y_1 \leftrightarrow  y_4$, $y_2 \leftrightarrow  y_5$ and $y_3 \leftrightarrow  y_6$, which permutes opposite nodes of the original quiver in \fref{quivers_dP3}.

\subsubsection{Factorized Partition Functions from Quiver Condensation}

As we have discussed in Section \ref{section_non_chiral_quivers}, non-chiral quivers give rise to factorized partition functions. This is because the node $k$ that is dualized at every step is of the form shown in \fref{non-chiral_node}.
Let us now introduce the idea of {\it quiver condensation}, which corresponds to identifying certain gauge groups in a quiver. More explicitly, condensing the quiver means that the $y_i$ variables for some of the quiver nodes are identified, giving rise to partially un-refined partition functions. Starting from a chiral quiver we can produce a non-chiral one by condensation. As a result, the associated un-refinement of the partition functions becomes factorized.

Let us illustrate these ideas with phase I of $dP_3$, in which we identify the original charges of opposite nodes in the quiver, i.e. if we set $y_1=y_4=a$, $y_2=y_5=b$ and $y_3=y_6=c$.  Under this identification, consecutive partition functions $Z_{2p}$ and $Z_{2p-1}$ and the recurrence equations \eref{y_transformation} and \eref{Z_transformation} simplify. In addition, partition functions factorize and we can provide closed expressions for them, i.e. we do not need to generate them recursively. Denoting $\mathcal{Z}_{p/2}\equiv Z_{2p}=Z_{2p-1},$ the partition functions are
\begin{eqnarray}
\mathcal{Z}_n & = & \prod_{i=0}^{n-1} (1+a^{i+1}b^{i+1}c^i)^{n-i} \prod_{i=0}^{n-1} (1+a^{i+1}b^{i}c^i)^{n-i} \nonumber \\
\mathcal{Z}_{n+1/2} & = & \prod_{i=0}^{n} (1+a^{i+1}b^{i}c^i)^{n-i+1} \prod_{i=0}^{n-1} (1+a^{i+1}b^{i+1}c^i)^{n-i}.
\label{Zs_dP3}
\end{eqnarray}
Indeed, it is straightforward to show that these partition functions are solutions of the recurrence equations, which simplify to
\beq
{\mathcal{Z}_{n+1/2} \, \mathcal{Z}_{n+2}\over \mathcal{Z}_{(n+1)+1/2} \, \mathcal{Z}_{(n+1)}}-1 = \mathfrak{q}_{n+1/2} \ \ \ \ \ \ \ \ \ \
{\mathcal{Z}_{n} \, \mathcal{Z}_{(n+1)+1/2}\over \mathcal{Z}_{(n+1)} \, \mathcal{Z}_{n+1/2}}-1 = \mathfrak{q}_n
\eeq
where we have renamed the prefactors as $\mathfrak{q}_{p/2}\equiv x_{2p}=x_{2p-1}$. In this notation, $\mathfrak{q}_n=a^{n+2}b^{n+1}c^{n+1}$ and 
$\mathfrak{q}_{n+1/2}=a^{n+2}b^{n+2}c^{n+1}$. Interestingly, the expressions in \eref{Zs_dP3} agree exactly with the partially refined partition functions derived in \cite{Cottrell:2010my}, which were obtained adapting the domino shuffling algorithm to the dimer model associated to phase I of $dP_3$.

\begin{figure}[h]
\begin{center}
\includegraphics[width=12cm]{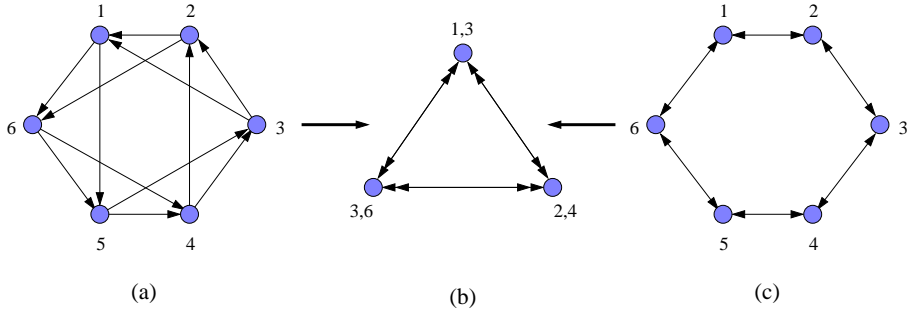}
\caption{The Model I of $dP_3$ quiver (a) and the $L^{3,3,3}$ quiver (c) result in the same quiver (b) upon condensation.}
\label{quiver_dP3_condensation}
\end{center}
\end{figure}

Different quiver theories can become equal by condensation. \fref{quiver_dP3_condensation} shows the example of $dP_3$ and $L^{3,3,3}.$ If we consider sequences of dualities that are identified by condensation, then the corresponding partially un-refined partition functions are equal for both theories.

\subsection{Identical Sequences of Partition Functions from Different Geometries: the $dP_2$ and $PdP_2$ Example}

\label{section_dP2_PdP2}

In this section we present two examples that illustrate how the same sequence of partition functions can arise from different toric geometries. In addition, the generated sequence is related to the Somos-5 sequence.

\subsection*{Del Pezzo 2}

\begin{figure}[h]
\begin{center}
\includegraphics[width=11.5cm]{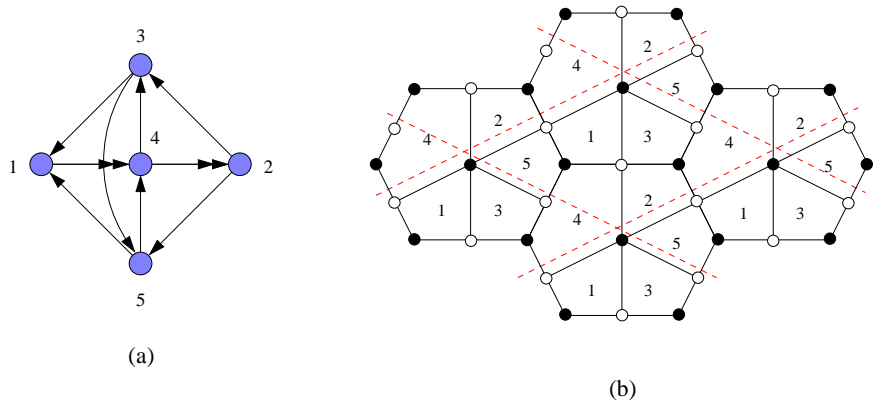}
\caption{a) Quiver and b) dimer model for phase II of $dP_2$.}
\label{quiver_dimer_dP2_II}
\end{center}
\end{figure}

Let us consider the cascade that only involves phase II of $dP_2$, which starts from the quiver in \fref{quivers_dP2} and corresponds to repeating the sequence of dualizations on nodes $(3,1,4,2,5)$. The toric diagram for $dP_2$ is given in \fref{fig:toric_dP2}. After each dualization, the gauge theory comes back to itself up to a permutation of its nodes. 
As in some of the previous examples, it is straightforward to give a closed expression for the prefactors in the cluster transformations in terms of a generating function. They are given by

\beq
x_n=\prod_{i=1}^5 y_i^{g_{s,(n-1)}}  \ \ \ \ \ g={1\over (1-q)(1-q^4)} 
\label{dP2_prefactor_1}
\eeq
Appendix \ref{section_partition_functions}, presents the explicit partition functions for the first steps in the cascade. Setting $y_i=1$, we see that the number of pyramid partitions corresponds to the Somos-5 sequence:
\beq
\text{{\bf Somos-5 sequence: }} 2, 3, 5, 11, 37, 83, 274, \ldots
\nonumber
\eeq

\noindent The Somos-5 sequence was also obtained in \cite{MR2317336}, by applying the octahedron recurrence to a graph that is different from the one in \fref{quiver_dimer_dP2_II}.b, i.e. a graph that can be interpreted as the dimer model associated to a different toric geometry. It is very illustrative to understand the connection between our results and those in \cite{MR2317336}, which we do below.

\subsection*{Pseudo del Pezzo 2}

\begin{figure}[h]
\begin{center}
\includegraphics[width=1.8cm]{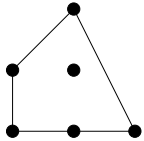}
\caption{Toric diagram for $PdP_2$.}
\label{toric_PdP2}
\end{center}
\end{figure}

\begin{figure}[h]
\begin{center}
\includegraphics[width=10cm]{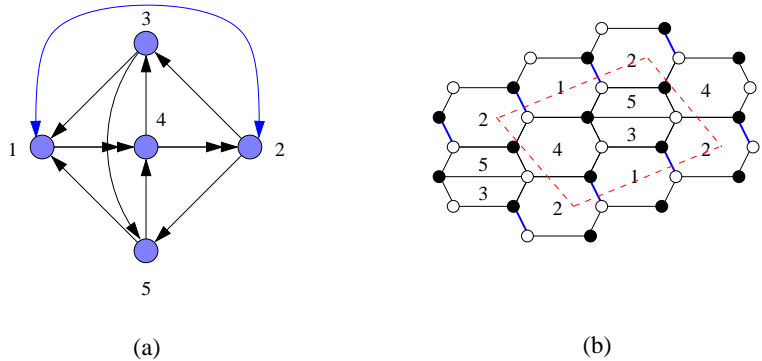}
\caption{a) Quiver diagram and b) dimer model for $PdP_2$.}
\label{quiver_dimer_PdP2}
\end{center}
\end{figure}

Let us consider the toric diagram in \fref{toric_PdP2}. 
This toric diagram has four corners. As a result the standard octahedron recurrence applies to this theory which implies that, unlike $dP_2$, it falls into the class of theories studied in \cite{MR2317336}. This geometry was denoted Pseudo del Pezzo 2 ($PdP_2$) in \cite{Feng:2004uq}, where its associated gauge theory was first introduced. \fref{quiver_dimer_PdP2} shows the gauge theory and dimer model for $PdP_2$. In fact, the dimer model in \fref{quiver_dimer_PdP2}.b is precisely the graph considered in \cite{MR2317336} in connection to the Somos-5 sequence.

The $PdP_2$ theory is a close cousin of phase II of $dP_2$, their matter content only differs by a vector pair of bifundamentals (i.e. a bidirectional arrow) connecting nodes 1 and 3, which we have indicated in blue in both the quiver and dimer model. It is important to stress that the edges associated to these extra fields are separated and hence there is no simple operation on the dimer that gives them a mass, removing them from the dimer and ending in the one for $dP_2$ shown in \fref{quiver_dimer_dP2_II}. A connection to $dP_2$ would be straightforward if, instead, the two extra edges that are present in $PdP_2$ were connected by a 2-valent node. This is because 2-valent nodes do not modify the perfect matching content and hence pyramid partitions are identical \cite{Franco:2005rj}.

$PdP_2$ has a cascade that follows the same sequence of dualities as the one discussed above for phase II of $dP_2$. Along this cascade, the prefactors and partition functions for $PdP_2$ are identical to those of $dP_2$. Explicit expressions for the first steps in the cascade are given in Appendix \ref{section_partition_functions}. The reason for this identification is that the dualized node never involves the bidirectional arrow, which distinguishes $dP_2$ from $PdP_2$, and effectively sees the same quiver.

While the partitions functions for both theories coincide along an infinite sequence of Seiberg dualities, it is natural to expect that they are distinct along other directions in the space of dual theories. In other words, while a specific low dimensional slice of their spaces of chambers agrees, their full higher dimensional structure is different.
We have seen that the Somos-5 sequence arises from pyramid partitions associated to both $PdP_2$ (which has a toric diagram with four and hence fits naturally into the analysis in \cite{MR2317336}) and $dP_2$ (which has a toric diagram with five corners).

\subsection{Stable Variables}

\label{section_stable_variables}

Up to now we have been writing partition functions in terms of the initial coefficients $y_i$, expressing the new coefficient variables $x_i$ after a sequence of Seiberg dualities as Laurent monomials in the original coefficients. 
Another natural basis for writing pyramid partitions is the one given by the brane charges. If we re-write the partition function as a function of the new coefficient variables, the resulting expressions display a surprising stabilization property. The first several terms of each partition function coincide. It is natural to conjecture that these polynomials will always converge as a formal power series expansion.

We illustrate this phenomenon for $dP_1$ below. We have replaced $x_i$ by its inverse to simplify the form of the partition functions. In order to facilitate comparison, it is convenient to present the partition functions in a table. The variables at different steps in the cascade differ in an obvious relabeling of the indices associated to the permutation of gauge groups that takes the quiver to its original form.

\begin{table}[htdp]
{\footnotesize
\begin{center}
\begin{tabular}{|c|c|c|c|c|}
\hline
\ \ \ \ \ $Z_1$ \ \ \ \ \ & \ \ \ \ \ $Z_2$ \ \ \ \ \ & \ \ \ \ \ $Z_3$ \ \ \ \ \ & \ \ \ \ \ $Z_4$ \ \ \ \ \ & \ \ \ \ \ $Z_5$ \ \ \ \ \ \\
\hline
1 & 1 & 1 & 1 & 1 \\
$x_1$ & 	$x_2$ 	& $x_3$		 & $x_4$ 					& $x_1$ \\ \hline
        & 	$x_1 x_2$ & $x_2 x_3$ 	& $x_3 x_4$ 				& $x_1 x_4$ \\ 
        &      &   $2 x_1 x_3^2$         	& $2 x_2 x_4^2$ 			& $2 x_1^2 x_3$\\
        &      & $x_1 x_2 x_3^3$       	& $x_2 x_3 x_4^3$ 			& $x_1^3 x_3 x_4$ \\
        &	&                                   	&                          			& $3 x_1^3 x_3^2$ \\ 
        &      & $ x_1^2 x_3^4$          	& $x_2^2 x_4^4$ 			&  \\
        &	&					& $3 x_1 x_3^2 x_4$ 		& $3 x_1 x_2 x_4^2 $ \\
        &	&					& $x_1 x_3^3 x_4$ 			& $x_1 x_2 x_4^3$ \\
        &	&					& $2 x_1 x_2 x_3^2 x_4^2$ 	& $2 x_1^2 x_2 x_3 x_4^2$\\
        &	&					& $3 x_1^2 x_3^4 x_4$ 		& $3 x_1 x_2^2 x_4^4$ \\
        &	&					&						& $2 x_1^4 x_3^2 x_4$ \\
        &	&					& 						& $2 x_1^5 x_3^3$ \\
        &	&					& $2 x_1 x_2 x_3^3 x_4^3$ 	& $2 x_1^3 x_2 x_3 x_4^3$ \\
        &	&					& $2 x_1 x_2^2 x_3^2 x_4^4$ 	& $2 x_1^4 x_2 x_3^2 x_4^2$\\ 
        &	&					& $x_1^3 x_3^6 x_4$ 		&\\
        &	&					& $x_1^2 x_2 x_3^5 x_4^3$	& $x_1^3 x_2^2 x_3 x_4^5$ \\
        &	&					& $x_1^2 x_2^2 x_3^4 x_4^4$	& $4 x_1^4 x_2^2 x_3^2 x_4^4$ \\
        &	&					&						& $2 x_1^3 x_2 x_3 x_4^3$ \\
        &	&					&						& $2 x_1^4 x_2 x_3^2 x_4^2$ \\
        &	&					&						& $\vdots$ \\
\hline
\end{tabular}
\end{center}}
\caption{Partition functions for $dP_1$ in terms of stable variables.}
\label{tab:default}
\end{table}

\bigskip

\section{Conclusions and Outlook}
\label{section_conclusion}

In this paper we have initiated a comprehensive investigation of the pyramids of finite and infinite type associated to general brane tilings. These tilings correspond to general, toric Calabi-Yau threefolds, including those with vanishing 4-cycles, which give rise to chiral quivers. We introduced various ways for defining and studying these pyramids based on quiver gauge theories and geometries. We also showed how cluster transformations provide an efficient tool for computing pyramid partition functions. There are several directions in which to extend the ideas of this paper. We mention some of them below.

We believe our framework will lead to a simple combinatorial proof of the cluster transformation properties of pyramid partition functions, by adapting the methods of \cite{MR2074946,MR2317336}.
It would be interesting to investigate the Donaldson-Thomas invariants that are obtained from mutations of non-toric nodes. Sequences of Seiberg dualities that include general, non-toric dualizations have been mapped using duality webs in \cite{hep-th/0110028,hep-th/0306092}. In addition, non-toric periodic cascades have been discussed for $dP_1$ in \cite{hep-th/0402120} and for $F_0$ in \cite{jeong1}. We hope our ideas will be useful for this purpose and that the connection to the multidimensional octahedron recurrence will allow a systematic study of far more general non-toric cascades.
 
Mirror symmetry relates, via the untwisting map, brane tilings on a torus to tilings of a Riemann surface in the mirror manifold \cite{hep-th/0511287}.  These tilings can be further refined to an ideal triangulation of the Riemann surface \cite{arXiv:1105.1777}.  For cluster algebras associated to triangulated surfaces, a combinatorial formula for the cluster variables in terms of perfect matchings was recently discovered \cite{MR2807089}.  We believe our combinatorial description of cluster variables associated to brane tilings should be closely related or even equivalent to
\cite{MR2807089}. It would be interesting to investigate the translation between the two perspectives.

The relationships between Seiberg dualities are encoded by the cluster modular groupoid \cite{MR2263192}.  The conjectural relationship between successive dualizations of nodes $\alpha$ and $\beta$ depends on the number of arrows between nodes $\alpha$ and $\beta$ through $\epsilon_{\alpha \beta} = \#arr(\alpha \rightarrow \beta) - \#arr(\beta \rightarrow \alpha).$
The conjectured pentagon relation states that if $\epsilon_{\alpha \beta} = -1,$ then the sequence of Seiberg dualities
$(\alpha, \beta, \alpha, \beta, \alpha)$ returns the quiver to its original form, up to a possible permutation of node labels.  It would be interesting to verify if this is true for general quiver gauge theories with superpotentials.

Recently, Goncharov and Kenyon discovered an exciting construction of integrable systems from brane tilings \cite{Goncharov:2011hp}.  This correspondence was further investigated in \cite{arXiv:1105.1777,arXiv:1107.1244}. Cluster transformations have interesting connections to integrable systems.
In \eqref{y_transformation} we saw that the cluster algebra coefficients transform as
$$
x_j' = 
\begin{cases}
x_k^{-1} & \text{if } j = k, \\
x_j \prod_{arr(k \rightarrow j)} x_k  & \text{if } j \neq k \text{ and } y_k \text{ has positive exponent,} \\
x_j \prod_{arr(j \rightarrow k)} x_k  & \text{if } j \neq k \text{ and } y_k \text{ has negative exponent.}
\end{cases}
$$
This transformation is a partial tropicalization of the transformation properties of the $w$ variables in the integrable system constructed from a brane tiling \cite{Goncharov:2011hp}.
$$
w_j' = 
\begin{cases}
w_k^{-1} & \text{if } j = k, \\
w_j \prod_{arr(k \rightarrow j)} (1 + w_k^{-1})^{-1} \prod_{arr(j \rightarrow k)} (1 + w_k)  & \text{if } j \neq k .
\end{cases}
$$
The discussion in Section \ref{section_duality_cascades} provides an efficient manner for generating periodic duality cascades.  Each period of the cascade corresponds to an auto-B\"acklund-Darboux transformation of the Goncharov-Kenyon integrable system \cite{Goncharov:2011hp,arXiv:1107.1244}. These transformations correspond to discrete time evolution of the integrable system. In retrospect, this is natural since the octahedron recurrence was first discovered as a discretization of the Hirota equation \cite{MR638804}. It would be interesting to continue exploring the connection between BPS partition functions and integrable systems from the viewpoint of the octahedron recurrence.

One of Speyer's motivations in \cite{MR2317336} for studying the octahedron recurrence was to give a simple combinatorial proof of the integrality of the Somos sequence. Somos sequences are non-linear recurrence relations that have many surprising integrality properties.  They were introduced by Somos to give a combinatorial analog of the addition formulas in the theory of elliptic functions.  We suggest that these addition laws can be thought of as the discrete time evolution of the integrable system associated to a brane tiling.  If true, this interpretation would close the circle of ideas inspired by the Somos sequence.  The ending of the Somos story is only the first chapter in a much larger story involving cluster algebras, Donaldson-Thomas invariants, and integrable systems.  
Cluster algebras have also appeared in connection with integrable systems associated to four dimensional $\mathcal{N} = 2$ gauge theories \cite{Cecotti:2010fi,Gaiotto:2010be}.  We expect this to be a fruitful area of research for years to come.

\bigskip

\section*{Acknowledgments}

We would like to thank Mauricio Romo for collaboration at the initial stages of this project. We also thank B. Young for correspondence and confirming agreement between our computations and the updated results of \cite{Cottrell:2010my}. The work of R. E. was supported in part by the National Science Foundation under grant DMS-1007414 and by the World Premier International Research Center Initiative (WPI Initiative), MEXT, Japan. R. E. and S. F. were supported by the National Science Foundation under Grant No. PHY05-51164. S. F. was also supported by the STFC.


\appendix

\newpage

\section{The $dP_2$ and $dP_3$ Gauge Theories}

\label{section_dP2_and_dP3}

In this appendix we summarize the toric phases for $dP_2$ and $dP_3$, which are studied in the body of the paper \cite{Feng:2002zw}.

\subsection*{Del Pezzo 2}

\begin{figure}[h]
\begin{center}
\includegraphics[width=9cm]{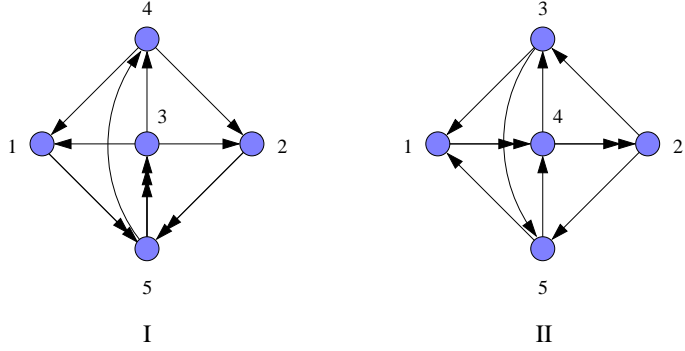}
\caption{Quiver diagrams for the two toric phases of $dP_2$.}
\label{quivers_dP2}
\end{center}
\end{figure}

There are two toric phases for $dP_2$, whose quivers are shown in \fref{quivers_dP2}. The corresponding superpotentials are

\beq
\begin{array}{ccl}
W_{I} & = & [X_{41}X_{15}X_{54} - X_{42}X_{25}X_{54}] - [X_{41}Y_{15}X_{53}X_{34} - X_{42}Y_{25}Y_{53}X_{34}] \\
& - & [X_{31}X_{15}Y_{53} - X_{32}X_{25}X_{53}] + [X_{31}Y_{15}Z_{53} - X_{32}Y_{25}Z_{53}] \\ \\
W_{II} & = & [X_{43}X_{35}X_{54}] - [X_{54}Y_{41}X_{15} + X_{43}X_{32}Y_{24}] \\
& + & [Y_{24}X_{41}X_{15}X_{52} + X_{32}X_{24}Y_{41}X_{13}] - [X_{24}X_{41}X_{13}X_{35}X_{52}]
\end{array}
\eeq

\subsection*{Del Pezzo 3}

\begin{figure}[h]
\begin{center}
\includegraphics[width=14cm]{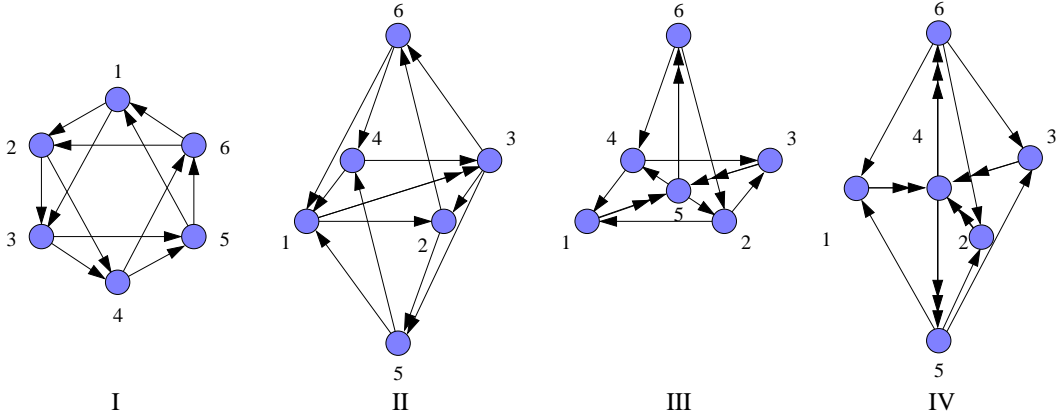}
\caption{Quiver diagrams for the four toric phases of $dP_3$.}
\label{quivers_dP3}
\end{center}
\end{figure}

There are two toric phases for $dP_3$, whose quivers are shown in \fref{quivers_dP3}. The corresponding superpotentials are

\beq
\begin{array}{ccl}
W_{I}& = & X_{12}X_{23}X_{34}X_{45}X_{56}X_{61}+[X_{13}X_{35}X_{51}+X_{24}X_{46}X_{62}] \\
& - & [X_{23}X_{35}X_{56}X_{62}-X_{13}X_{34}X_{46}X_{61}-X_{12}X_{24}X_{45}X_{51}] \\ \\
W_{II} & = & [X_{12}X_{26}X_{61} - X_{12}X_{25}X_{51} + X_{36}X_{64}X_{43} - X_{35}X_{54}X_{43}] \\
&+&[-X_{61}X_{13}X_{36} + X_{51}Y_{13}X_{35}] + [-X_{26}X_{64}X_{41}Y_{13}X_{32} + X_{25}X_{54}X_{41}X_{13}X_{32}]
\end{array}
\nonumber
\eeq

\beq
\begin{array}{ccl}
W_{III} & = & [X_{41}X_{15}X_{54} - X_{54}X_{43}X_{35} + Y_{35}X_{52}X_{23} - X_{52}X_{21}Y_{15}] \\
& + & [-X_{41}Y_{15}X_{56}X_{64} + X_{64}X_{43}Y_{35}Y_{56} - X_{23}X_{35}X_{56}X_{62} + X_{62}X_{21}X_{15}Y_{56}] \\ \\
W_{IV} & = & [X_{41}X_{16}X_{64} + X_{43}X_{36}Y_{64} + X_{42}X_{26}Z_{64}] - [X_{41}Y_{16}Y_{64} + X_{43}Y_{36}Z_{64} + X_{42}Y_{26}X_{64}] \\
& + & [X_{51}Y_{16}X_{65} + X_{53}Y_{36}Y_{65} + X_{52}Y_{26}Z_{65}]
- [X_{51}X_{16}Y_{65} + X_{53}X_{36}Z_{65} + X_{52}X_{26}X_{65}]
\end{array}
\eeq

\section{A Sampler of Partition Functions}

\label{section_partition_functions}

Here we present explicit partition functions for some of the models discussed in Section \ref{section_explicit_examples}. These examples are intended both as reference and as an illustration of the kind of expressions that are straightforwardly generated by the cluster transformations. Alternatively, these partition functions can be directly obtained by first determining the stones in the pyramid and then finding its partitions, with the help of a computer if necessary, as discussed in \cite{Mozgovoy:2008fd}. Even the first step becomes prohibitively involved for large pyramids. It is important to emphasize that the partition functions we present below were exclusively calculated using the cluster transformations, without actually constructing the corresponding pyramids. Of course, checking the simplest examples against direct computation is straightforward.

\bigskip

\subsection*{Del Pezzo 1}

{\footnotesize
\beq
\begin{array}{rcl}
Z_1 & = & 1+y_1 \\

Z_2 & = & 1+y_2+y_1 y_2 \\

Z_3 & = & 1+2 y_1+y_1^2+y_1^2 y_3+y_1^2 y_2 y_3+y_1^3 y_2 y_3 \\

Z_4 & = & 1+2 y_2+2 y_1 y_2+y_2^2+2 y_1 y_2^2+y_1^2 y_2^2+y_2^2 y_4+3 y_1 y_2^2 y_4+3 y_1^2 y_2^2 y_4+y_1^3 y_2^2
   y_4+y_1^2 y_2^2 y_3 y_4+y_1^3 y_2^2 y_3 y_4\\ & & +y_1^2 y_2^3 y_3 y_4+2 y_1^3 y_2^3 y_3 y_4+y_1^4 y_2^3
   y_3 y_4 \\

Z_5 & = & 1+3 y_1+3 y_1^2+y_1^3+2 y_1^2 y_3+2 y_1^3 y_3+2 y_1^2 y_2 y_3+4 y_1^3 y_2 y_3+2 y_1^4 y_2 y_3+y_1^3
   y_3^2+3 y_1^3 y_2 y_3^2+2 y_1^4 y_2 y_3^2+3 y_1^3 y_2^2 y_3^2\\ & & +4 y_1^4 y_2^2 y_3^2+y_1^5 y_2^2
   y_3^2+y_1^3 y_2^3 y_3^2+2 y_1^4 y_2^3 y_3^2+y_1^5 y_2^3 y_3^2+y_1^3 y_2^2 y_3^2 y_4+2 y_1^4 y_2^2
   y_3^2 y_4+y_1^5 y_2^2 y_3^2 y_4+y_1^3 y_2^3 y_3^2 y_4+3 y_1^4 y_2^3 y_3^2 y_4\\ & & +3 y_1^5 y_2^3 y_3^2
   y_4+y_1^6 y_2^3 y_3^2 y_4+y_1^5 y_2^2 y_3^3 y_4+2 y_1^5 y_2^3 y_3^3 y_4+2 y_1^6 y_2^3 y_3^3
   y_4+y_1^5 y_2^4 y_3^3 y_4+2 y_1^6 y_2^4 y_3^3 y_4+y_1^7 y_2^4 y_3^3 y_4 \\

Z_6 & = & 1+3 y_2+3 y_1 y_2+3 y_2^2+6 y_1 y_2^2+3 y_1^2 y_2^2+y_2^3+3 y_1 y_2^3+3 y_1^2 y_2^3+y_1^3 y_2^3+2 y_2^2
   y_4+6 y_1 y_2^2 y_4+6 y_1^2 y_2^2 y_4+2 y_1^3 y_2^2 y_4\\ & & +2 y_2^3 y_4+8 y_1 y_2^3 y_4+12 y_1^2 y_2^3
   y_4+8 y_1^3 y_2^3 y_4+2 y_1^4 y_2^3 y_4+2 y_1^2 y_2^2 y_3 y_4+2 y_1^3 y_2^2 y_3 y_4+4 y_1^2 y_2^3
   y_3 y_4+8 y_1^3 y_2^3 y_3 y_4\\ & & +4 y_1^4 y_2^3 y_3 y_4+2 y_1^2 y_2^4 y_3 y_4+6 y_1^3 y_2^4 y_3 y_4+6
   y_1^4 y_2^4 y_3 y_4+2 y_1^5 y_2^4 y_3 y_4+y_2^3 y_4^2+5 y_1 y_2^3 y_4^2+10 y_1^2 y_2^3 y_4^2+10
   y_1^3 y_2^3 y_4^2\\ & & +5 y_1^4 y_2^3 y_4^2+y_1^5 y_2^3 y_4^2+3 y_1^2 y_2^3 y_3 y_4^2+9 y_1^3 y_2^3 y_3
   y_4^2+9 y_1^4 y_2^3 y_3 y_4^2+3 y_1^5 y_2^3 y_3 y_4^2+2 y_1^2 y_2^4 y_3 y_4^2+8 y_1^3 y_2^4 y_3
   y_4^2+12 y_1^4 y_2^4 y_3 y_4^2\\ & & +8 y_1^5 y_2^4 y_3 y_4^2+2 y_1^6 y_2^4 y_3 y_4^2+y_1^3 y_2^3 y_3^2
   y_4^2+4 y_1^4 y_2^3 y_3^2 y_4^2+3 y_1^5 y_2^3 y_3^2 y_4^2+2 y_1^3 y_2^4 y_3^2 y_4^2+8 y_1^4 y_2^4
   y_3^2 y_4^2+10 y_1^5 y_2^4 y_3^2 y_4^2\\ & & +4 y_1^6 y_2^4 y_3^2 y_4^2+y_1^3 y_2^5 y_3^2 y_4^2+4 y_1^4
   y_2^5 y_3^2 y_4^2+6 y_1^5 y_2^5 y_3^2 y_4^2+4 y_1^6 y_2^5 y_3^2 y_4^2+y_1^7 y_2^5 y_3^2 y_4^2+y_1^5
   y_2^3 y_3^3 y_4^2+3 y_1^5 y_2^4 y_3^3 y_4^2\\ & & +2 y_1^6 y_2^4 y_3^3 y_4^2+3 y_1^5 y_2^5 y_3^3 y_4^2+4
   y_1^6 y_2^5 y_3^3 y_4^2+y_1^7 y_2^5 y_3^3 y_4^2+y_1^5 y_2^6 y_3^3 y_4^2+2 y_1^6 y_2^6 y_3^3
   y_4^2+y_1^7 y_2^6 y_3^3 y_4^2+y_1^3 y_2^5 y_3^2 y_4^3\\ & & +4 y_1^4 y_2^5 y_3^2 y_4^3+6 y_1^5 y_2^5 y_3^2
   y_4^3+4 y_1^6 y_2^5 y_3^2 y_4^3+y_1^7 y_2^5 y_3^2 y_4^3+2 y_1^5 y_2^5 y_3^3 y_4^3+4 y_1^6 y_2^5
   y_3^3 y_4^3+2 y_1^7 y_2^5 y_3^3 y_4^3+2 y_1^5 y_2^6 y_3^3 y_4^3\\ & & +6 y_1^6 y_2^6 y_3^3 y_4^3+6 y_1^7
   y_2^6 y_3^3 y_4^3+2 y_1^8 y_2^6 y_3^3 y_4^3+y_1^7 y_2^5 y_3^4 y_4^3+2 y_1^7 y_2^6 y_3^4 y_4^3+2
   y_1^8 y_2^6 y_3^4 y_4^3+y_1^7 y_2^7 y_3^4 y_4^3+2 y_1^8 y_2^7 y_3^4 y_4^3\\ & & +y_1^9 y_2^7 y_3^4 y_4^3 \\

Z_7 & = & 1+4 y_1+6 y_1^2+4 y_1^3+y_1^4+3 y_1^2 y_3+6 y_1^3 y_3+3 y_1^4 y_3+3 y_1^2 y_2 y_3+9 y_1^3 y_2 y_3+9
   y_1^4 y_2 y_3+3 y_1^5 y_2 y_3+2 y_1^3 y_3^2+3 y_1^4 y_3^2\\ & & +6 y_1^3 y_2 y_3^2+12 y_1^4 y_2 y_3^2+6
   y_1^5 y_2 y_3^2+6 y_1^3 y_2^2 y_3^2+15 y_1^4 y_2^2 y_3^2+12 y_1^5 y_2^2 y_3^2+3 y_1^6 y_2^2 y_3^2+2
   y_1^3 y_2^3 y_3^2+6 y_1^4 y_2^3 y_3^2+6 y_1^5 y_2^3 y_3^2\\ & & +2 y_1^6 y_2^3 y_3^2+y_1^4 y_3^3+5 y_1^4
   y_2 y_3^3+3 y_1^5 y_2 y_3^3+10 y_1^4 y_2^2 y_3^3+12 y_1^5 y_2^2 y_3^3+3 y_1^6 y_2^2 y_3^3+10 y_1^4
   y_2^3 y_3^3+18 y_1^5 y_2^3 y_3^3+9 y_1^6 y_2^3 y_3^3\\ & & +y_1^7 y_2^3 y_3^3+5 y_1^4 y_2^4 y_3^3+12 y_1^5
   y_2^4 y_3^3+9 y_1^6 y_2^4 y_3^3+2 y_1^7 y_2^4 y_3^3+y_1^4 y_2^5 y_3^3+3 y_1^5 y_2^5 y_3^3+3 y_1^6
   y_2^5 y_3^3+y_1^7 y_2^5 y_3^3+2 y_1^3 y_2^2 y_3^2 y_4\\ & & +6 y_1^4 y_2^2 y_3^2 y_4+6 y_1^5 y_2^2 y_3^2
   y_4+2 y_1^6 y_2^2 y_3^2 y_4+2 y_1^3 y_2^3 y_3^2 y_4+8 y_1^4 y_2^3 y_3^2 y_4+12 y_1^5 y_2^3 y_3^2
   y_4+8 y_1^6 y_2^3 y_3^2 y_4+2 y_1^7 y_2^3 y_3^2 y_4\\ & & +3 y_1^4 y_2^2 y_3^3 y_4+7 y_1^5 y_2^2 y_3^3
   y_4+4 y_1^6 y_2^2 y_3^3 y_4+9 y_1^4 y_2^3 y_3^3 y_4+26 y_1^5 y_2^3 y_3^3 y_4+25 y_1^6 y_2^3 y_3^3
   y_4+8 y_1^7 y_2^3 y_3^3 y_4+9 y_1^4 y_2^4 y_3^3 y_4\\ & & +31 y_1^5 y_2^4 y_3^3 y_4+39 y_1^6 y_2^4 y_3^3
   y_4+21 y_1^7 y_2^4 y_3^3 y_4+4 y_1^8 y_2^4 y_3^3 y_4+3 y_1^4 y_2^5 y_3^3 y_4+12 y_1^5 y_2^5 y_3^3
   y_4+18 y_1^6 y_2^5 y_3^3 y_4+12 y_1^7 y_2^5 y_3^3 y_4\\ & & +3 y_1^8 y_2^5 y_3^3 y_4+2 y_1^6 y_2^2 y_3^4
   y_4+8 y_1^6 y_2^3 y_3^4 y_4+6 y_1^7 y_2^3 y_3^4 y_4+12 y_1^6 y_2^4 y_3^4 y_4+18 y_1^7 y_2^4 y_3^4
   y_4+6 y_1^8 y_2^4 y_3^4 y_4+8 y_1^6 y_2^5 y_3^4 y_4\\ & & +18 y_1^7 y_2^5 y_3^4 y_4+12 y_1^8 y_2^5 y_3^4
   y_4+2 y_1^9 y_2^5 y_3^4 y_4+2 y_1^6 y_2^6 y_3^4 y_4+6 y_1^7 y_2^6 y_3^4 y_4+6 y_1^8 y_2^6 y_3^4
   y_4+2 y_1^9 y_2^6 y_3^4 y_4+y_1^4 y_2^3 y_3^3 y_4^2\\ & & +3 y_1^5 y_2^3 y_3^3 y_4^2+3 y_1^6 y_2^3 y_3^3
   y_4^2+y_1^7 y_2^3 y_3^3 y_4^2+4 y_1^4 y_2^4 y_3^3 y_4^2+16 y_1^5 y_2^4 y_3^3 y_4^2+24 y_1^6 y_2^4
   y_3^3 y_4^2+16 y_1^7 y_2^4 y_3^3 y_4^2+4 y_1^8 y_2^4 y_3^3 y_4^2\\ & & +3 y_1^4 y_2^5 y_3^3 y_4^2+15 y_1^5
   y_2^5 y_3^3 y_4^2+30 y_1^6 y_2^5 y_3^3 y_4^2+30 y_1^7 y_2^5 y_3^3 y_4^2+15 y_1^8 y_2^5 y_3^3 y_4^2+3
   y_1^9 y_2^5 y_3^3 y_4^2+2 y_1^6 y_2^3 y_3^4 y_4^2+2 y_1^7 y_2^3 y_3^4 y_4^2\\ & & +8 y_1^6 y_2^4 y_3^4
   y_4^2+16 y_1^7 y_2^4 y_3^4 y_4^2+8 y_1^8 y_2^4 y_3^4 y_4^2+10 y_1^6 y_2^5 y_3^4 y_4^2+30 y_1^7 y_2^5
   y_3^4 y_4^2+30 y_1^8 y_2^5 y_3^4 y_4^2+10 y_1^9 y_2^5 y_3^4 y_4^2+4 y_1^6 y_2^6 y_3^4 y_4^2\\ & & +16 y_1^7
   y_2^6 y_3^4 y_4^2+24 y_1^8 y_2^6 y_3^4 y_4^2+16 y_1^9 y_2^6 y_3^4 y_4^2+4 y_1^{10} y_2^6 y_3^4
   y_4^2+y_1^7 y_2^3 y_3^5 y_4^2+4 y_1^7 y_2^4 y_3^5 y_4^2+4 y_1^8 y_2^4 y_3^5 y_4^2+6 y_1^7 y_2^5
   y_3^5 y_4^2\\ & & +12 y_1^8 y_2^5 y_3^5 y_4^2+6 y_1^9 y_2^5 y_3^5 y_4^2+4 y_1^7 y_2^6 y_3^5 y_4^2+12 y_1^8
   y_2^6 y_3^5 y_4^2+12 y_1^9 y_2^6 y_3^5 y_4^2+4 y_1^{10} y_2^6 y_3^5 y_4^2+y_1^7 y_2^7 y_3^5 y_4^2+4
   y_1^8 y_2^7 y_3^5 y_4^2\\ & & +6 y_1^9 y_2^7 y_3^5 y_4^2+4 y_1^{10} y_2^7 y_3^5 y_4^2+y_1^{11} y_2^7 y_3^5
   y_4^2+y_1^4 y_2^5 y_3^3 y_4^3+6 y_1^5 y_2^5 y_3^3 y_4^3+15 y_1^6 y_2^5 y_3^3 y_4^3+20 y_1^7 y_2^5
   y_3^3 y_4^3+15 y_1^8 y_2^5 y_3^3 y_4^3\\ & & +6 y_1^9 y_2^5 y_3^3 y_4^3+y_1^{10} y_2^5 y_3^3 y_4^3+3 y_1^6
   y_2^5 y_3^4 y_4^3+12 y_1^7 y_2^5 y_3^4 y_4^3+18 y_1^8 y_2^5 y_3^4 y_4^3+12 y_1^9 y_2^5 y_3^4 y_4^3+3
   y_1^{10} y_2^5 y_3^4 y_4^3+2 y_1^6 y_2^6 y_3^4 y_4^3\\ & & +10 y_1^7 y_2^6 y_3^4 y_4^3+20 y_1^8 y_2^6 y_3^4
   y_4^3+20 y_1^9 y_2^6 y_3^4 y_4^3+10 y_1^{10} y_2^6 y_3^4 y_4^3+2 y_1^{11} y_2^6 y_3^4 y_4^3+2 y_1^7
   y_2^5 y_3^5 y_4^3+7 y_1^8 y_2^5 y_3^5 y_4^3+8 y_1^9 y_2^5 y_3^5 y_4^3\\ & & +3 y_1^{10} y_2^5 y_3^5 y_4^3+4
   y_1^7 y_2^6 y_3^5 y_4^3+16 y_1^8 y_2^6 y_3^5 y_4^3+24 y_1^9 y_2^6 y_3^5 y_4^3+16 y_1^{10} y_2^6
   y_3^5 y_4^3+4 y_1^{11} y_2^6 y_3^5 y_4^3+2 y_1^7 y_2^7 y_3^5 y_4^3+9 y_1^8 y_2^7 y_3^5 y_4^3\\ & & +16
   y_1^9 y_2^7 y_3^5 y_4^3+14 y_1^{10} y_2^7 y_3^5 y_4^3+6 y_1^{11} y_2^7 y_3^5 y_4^3+y_1^{12} y_2^7
   y_3^5 y_4^3+2 y_1^9 y_2^5 y_3^6 y_4^3+y_1^{10} y_2^5 y_3^6 y_4^3+6 y_1^9 y_2^6 y_3^6 y_4^3+8
   y_1^{10} y_2^6 y_3^6 y_4^3\\ & & +2 y_1^{11} y_2^6 y_3^6 y_4^3+6 y_1^9 y_2^7 y_3^6 y_4^3+13 y_1^{10} y_2^7
   y_3^6 y_4^3+8 y_1^{11} y_2^7 y_3^6 y_4^3+y_1^{12} y_2^7 y_3^6 y_4^3+2 y_1^9 y_2^8 y_3^6 y_4^3+6
   y_1^{10} y_2^8 y_3^6 y_4^3+6 y_1^{11} y_2^8 y_3^6 y_4^3\\ & & +2 y_1^{12} y_2^8 y_3^6 y_4^3+y_1^7 y_2^7
   y_3^5 y_4^4+5 y_1^8 y_2^7 y_3^5 y_4^4+10 y_1^9 y_2^7 y_3^5 y_4^4+10 y_1^{10} y_2^7 y_3^5 y_4^4+5
   y_1^{11} y_2^7 y_3^5 y_4^4+y_1^{12} y_2^7 y_3^5 y_4^4+2 y_1^9 y_2^7 y_3^6 y_4^4\\ & & +6 y_1^{10} y_2^7
   y_3^6 y_4^4+6 y_1^{11} y_2^7 y_3^6 y_4^4+2 y_1^{12} y_2^7 y_3^6 y_4^4+2 y_1^9 y_2^8 y_3^6 y_4^4+8
   y_1^{10} y_2^8 y_3^6 y_4^4+12 y_1^{11} y_2^8 y_3^6 y_4^4+8 y_1^{12} y_2^8 y_3^6 y_4^4+2 y_1^{13}
   y_2^8 y_3^6 y_4^4\\ & & +y_1^{11} y_2^7 y_3^7 y_4^4+y_1^{12} y_2^7 y_3^7 y_4^4+2 y_1^{11} y_2^8 y_3^7
   y_4^4+4 y_1^{12} y_2^8 y_3^7 y_4^4+2 y_1^{13} y_2^8 y_3^7 y_4^4+y_1^{11} y_2^9 y_3^7 y_4^4+3
   y_1^{12} y_2^9 y_3^7 y_4^4+3 y_1^{13} y_2^9 y_3^7 y_4^4\\ & & +y_1^{14} y_2^9 y_3^7 y_4^4
\end{array}
\eeq
}

\subsection*{Del Pezzo 2}

{\footnotesize
\beq
\begin{array}{rcl}
Z_1 & = & 1 + y_3 \\
Z_2 & = & 1 + y_3 + y_1 y_3 \\
Z_3 & = & 1 + y_3 + y_1 y_3 + y_1 y_3 y_4 + y_1 y_3^2 y_4 \\
Z_4 & = & 1 + y_3 + y_1 y_3 + y_1 y_3 y_4 + y_1 y_2 y_3 y_4 + y_1 y_3^2 y_4 + 2 y_1 y_2 y_3^2 y_4 + y_1^2 y_2 y_3^2 y_4 + y_1 y_2 y_3^3 y_4 + y_1^2 y_2 y_3^3 y_4 \\
Z_5 & = & 1 + 2 y_3 + y_1 y_3 + y_3^2 + y_1 y_3^2 + y_1 y_3 y_4 + y_1 y_2 y_3 y_4 + 
   2 y_1 y_3^2 y_4 + 3 y_1 y_2 y_3^2 y_4 + y_1^2 y_2 y_3^2 y_4 + y_1 y_3^3 y_4 \\ & & + 
   3 y_1 y_2 y_3^3 y_4 + 2 y_1^2 y_2 y_3^3 y_4 + y_1 y_2 y_3^4 y_4 + 
   y_1^2 y_2 y_3^4 y_4 + y_1 y_2 y_3^2 y_4 y_5 + 2 y_1 y_2 y_3^3 y_4 y_5 + 
   2 y_1^2 y_2 y_3^3 y_4 y_5 \\
 & & + y_1 y_2 y_3^4 y_4 y_5 + 2 y_1^2 y_2 y_3^4 y_4 y_5 + 
   y_1^3 y_2 y_3^4 y_4 y_5 + y_1^2 y_2 y_3^3 y_4^2 y_5 + 
   2 y_1^2 y_2 y_3^4 y_4^2 y_5 + y_1^3 y_2 y_3^4 y_4^2 y_5 + 
   y_1^2 y_2 y_3^5 y_4^2 y_5 \\ & & + y_1^3 y_2 y_3^5 y_4^2 y_5
\end{array}
\eeq
}
As explained in Section \ref{section_dP2_PdP2}, these partition functions are identical to the ones for $PdP_2$ along a specific sequence of Seiberg dualities. The reason for this identification is that, for this sequence, the dualized node never involves the bidirectional arrow, which distinguishes $dP_2$ from $PdP_2$.

\bigskip

\subsection*{Del Pezzo 3}

{\footnotesize
\beq
\begin{array}{rcl}
Z_1 & = & 1+y_1 \\

Z_2 & = & 1+y_4 \\

Z_3 & = & 1+y_1+y_1 y_2+y_1 y_2 y_4 \\

Z_4 & = & 1+y_4+y_4 y_5+y_1 y_4 y_5 \\

Z_5 & = & 1+2 y_1+y_1^2+y_1 y_2+y_1^2 y_2+y_1^2 y_2 y_3+y_1 y_2 y_4+y_1^2 y_2 y_4+2 y_1^2 y_2 y_3 y_4+y_1^2 y_2
   y_3 y_4^2+y_1^2 y_2 y_3 y_4 y_5+y_1^3 y_2 y_3 y_4 y_5 \\ & & +y_1^2 y_2 y_3 y_4^2 y_5+y_1^3 y_2 y_3 y_4^2
   y_5 \\

Z_6 & = & 1+2 y_4+y_4^2+y_4 y_5+y_1 y_4 y_5+y_4^2 y_5+y_1 y_4^2 y_5+y_4^2 y_5 y_6+2 y_1 y_4^2 y_5 y_6+y_1^2 y_4^2
   y_5 y_6+y_1 y_2 y_4^2 y_5 y_6+y_1^2 y_2 y_4^2 y_5 y_6 \\ & & +y_1 y_2 y_4^3 y_5 y_6+y_1^2 y_2 y_4^3 y_5 y_6 \\

Z_7 & = & 1+2 y_4+y_4^2+2 y_4 y_5+2 y_1 y_4 y_5+2 y_4^2 y_5+2 y_1 y_4^2 y_5+y_4^2 y_5^2+2 y_1 y_4^2 y_5^2+y_1^2
   y_4^2 y_5^2+y_4^2 y_5 y_6+2 y_1 y_4^2 y_5 y_6 \\ & & +y_1^2 y_4^2 y_5 y_6 +y_1 y_2 y_4^2 y_5 y_6+y_1^2 y_2
   y_4^2 y_5 y_6+y_1 y_2 y_4^3 y_5 y_6+y_1^2 y_2 y_4^3 y_5 y_6+y_4^2 y_5^2 y_6+3 y_1 y_4^2 y_5^2 y_6+3
   y_1^2 y_4^2 y_5^2 y_6 \\ & & +y_1^3 y_4^2 y_5^2 y_6+2 y_1 y_2 y_4^2 y_5^2 y_6+4 y_1^2 y_2 y_4^2 y_5^2 y_6+2
   y_1^3 y_2 y_4^2 y_5^2 y_6+y_1^2 y_2^2 y_4^2 y_5^2 y_6+y_1^3 y_2^2 y_4^2 y_5^2 y_6+y_1^2 y_2 y_3 y_4^2
   y_5^2 y_6 \\ & & +y_1^3 y_2 y_3 y_4^2 y_5^2 y_6+y_1^3 y_2^2 y_3 y_4^2 y_5^2 y_6+y_1 y_2 y_4^3 y_5^2 y_6+2
   y_1^2 y_2 y_4^3 y_5^2 y_6+y_1^3 y_2 y_4^3 y_5^2 y_6+y_1^2 y_2^2 y_4^3 y_5^2 y_6+y_1^3 y_2^2 y_4^3
   y_5^2 y_6 \\ & & +y_1^2 y_2 y_3 y_4^3 y_5^2 y_6+y_1^3 y_2 y_3 y_4^3 y_5^2 y_6+2 y_1^3 y_2^2 y_3 y_4^3 y_5^2
   y_6+y_1^3 y_2^2 y_3 y_4^4 y_5^2 y_6+y_1^2 y_2 y_3 y_4^3 y_5^3 y_6+2 y_1^3 y_2 y_3 y_4^3 y_5^3
   y_6 \\ & & +y_1^4 y_2 y_3 y_4^3 y_5^3 y_6+y_1^3 y_2^2 y_3 y_4^3 y_5^3 y_6+y_1^4 y_2^2 y_3 y_4^3 y_5^3
   y_6+y_1^3 y_2^2 y_3 y_4^4 y_5^3 y_6+y_1^4 y_2^2 y_3 y_4^4 y_5^3 y_6 \\

Z_8 & = & 1+2 y_1+y_1^2+2 y_1 y_2+2 y_1^2 y_2+y_1^2 y_2^2+y_1^2 y_2 y_3+y_1^2 y_2^2 y_3+2 y_1 y_2 y_4+2 y_1^2 y_2
   y_4+2 y_1^2 y_2^2 y_4+2 y_1^2 y_2 y_3 y_4 \\ & & +3 y_1^2 y_2^2 y_3 y_4+y_1^2 y_2^2 y_4^2+y_1^2 y_2 y_3
   y_4^2+3 y_1^2 y_2^2 y_3 y_4^2+y_1^2 y_2^2 y_3 y_4^3+y_1^2 y_2 y_3 y_4 y_5+y_1^3 y_2 y_3 y_4 y_5+2
   y_1^2 y_2^2 y_3 y_4 y_5 \\ & & +y_1^3 y_2^2 y_3 y_4 y_5+y_1^2 y_2 y_3 y_4^2 y_5+y_1^3 y_2 y_3 y_4^2 y_5+4
   y_1^2 y_2^2 y_3 y_4^2 y_5+2 y_1^3 y_2^2 y_3 y_4^2 y_5+2 y_1^2 y_2^2 y_3 y_4^3 y_5+y_1^3 y_2^2 y_3
   y_4^3 y_5 \\ & & +y_1^2 y_2^2 y_3 y_4^2 y_5^2+y_1^3 y_2^2 y_3 y_4^2 y_5^2+y_1^2 y_2^2 y_3 y_4^3 y_5^2+y_1^3
   y_2^2 y_3 y_4^3 y_5^2+y_1^2 y_2^2 y_3 y_4^2 y_5 y_6+y_1^3 y_2^2 y_3 y_4^2 y_5 y_6+y_1^3 y_2^3 y_3
   y_4^2 y_5 y_6 \\ & & +y_1^2 y_2^2 y_3 y_4^3 y_5 y_6+y_1^3 y_2^2 y_3 y_4^3 y_5 y_6+2 y_1^3 y_2^3 y_3 y_4^3 y_5
   y_6+y_1^3 y_2^3 y_3 y_4^4 y_5 y_6+y_1^2 y_2^2 y_3 y_4^3 y_5^2 y_6+2 y_1^3 y_2^2 y_3 y_4^3 y_5^2
   y_6 \\ & & +y_1^4 y_2^2 y_3 y_4^3 y_5^2 y_6+y_1^3 y_2^3 y_3 y_4^3 y_5^2 y_6+y_1^4 y_2^3 y_3 y_4^3 y_5^2
   y_6+y_1^3 y_2^3 y_3 y_4^4 y_5^2 y_6+y_1^4 y_2^3 y_3 y_4^4 y_5^2 y_6 \\

Z_9 & = & 1+3 y_4+3 y_4^2+y_4^3+2 y_4 y_5+2 y_1 y_4 y_5+4 y_4^2 y_5+4 y_1 y_4^2 y_5+2 y_4^3 y_5+2 y_1 y_4^3
   y_5+y_4^2 y_5^2+2 y_1 y_4^2 y_5^2+y_1^2 y_4^2 y_5^2+y_4^3 y_5^2 \\ & & +2 y_1 y_4^3 y_5^2+y_1^2 y_4^3 y_5^2+2
   y_4^2 y_5 y_6+4 y_1 y_4^2 y_5 y_6+2 y_1^2 y_4^2 y_5 y_6+2 y_1 y_2 y_4^2 y_5 y_6+2 y_1^2 y_2 y_4^2 y_5
   y_6+2 y_4^3 y_5 y_6+4 y_1 y_4^3 y_5 y_6 \\ & & +2 y_1^2 y_4^3 y_5 y_6+4 y_1 y_2 y_4^3 y_5 y_6+4 y_1^2 y_2
   y_4^3 y_5 y_6+2 y_1 y_2 y_4^4 y_5 y_6+2 y_1^2 y_2 y_4^4 y_5 y_6+y_4^2 y_5^2 y_6+3 y_1 y_4^2 y_5^2
   y_6+3 y_1^2 y_4^2 y_5^2 y_6 \\ & & +y_1^3 y_4^2 y_5^2 y_6+2 y_1 y_2 y_4^2 y_5^2 y_6+4 y_1^2 y_2 y_4^2 y_5^2
   y_6+2 y_1^3 y_2 y_4^2 y_5^2 y_6+y_1^2 y_2^2 y_4^2 y_5^2 y_6+y_1^3 y_2^2 y_4^2 y_5^2 y_6+y_1^2 y_2 y_3
   y_4^2 y_5^2 y_6 \\ \end{array}
\nonumber
\eeq

\beq
\begin{array}{rcl}
& & +y_1^3 y_2 y_3 y_4^2 y_5^2 y_6+y_1^3 y_2^2 y_3 y_4^2 y_5^2 y_6+2 y_4^3 y_5^2 y_6+6 y_1
   y_4^3 y_5^2 y_6+6 y_1^2 y_4^3 y_5^2 y_6+2 y_1^3 y_4^3 y_5^2 y_6+4 y_1 y_2 y_4^3 y_5^2 y_6+8 y_1^2 y_2
   y_4^3 y_5^2 y_6 \\ 
& & +4 y_1^3 y_2 y_4^3 y_5^2 y_6+2 y_1^2 y_2^2 y_4^3 y_5^2 y_6+2 y_1^3 y_2^2 y_4^3 y_5^2
   y_6+2 y_1^2 y_2 y_3 y_4^3 y_5^2 y_6+2 y_1^3 y_2 y_3 y_4^3 y_5^2 y_6+3 y_1^3 y_2^2 y_3 y_4^3 y_5^2
   y_6+2 y_1 y_2 y_4^4 y_5^2 y_6 \\ & & +4 y_1^2 y_2 y_4^4 y_5^2 y_6+2 y_1^3 y_2 y_4^4 y_5^2 y_6+y_1^2 y_2^2
   y_4^4 y_5^2 y_6+y_1^3 y_2^2 y_4^4 y_5^2 y_6+y_1^2 y_2 y_3 y_4^4 y_5^2 y_6+y_1^3 y_2 y_3 y_4^4 y_5^2
   y_6+3 y_1^3 y_2^2 y_3 y_4^4 y_5^2 y_6 \\ & & +y_1^3 y_2^2 y_3 y_4^5 y_5^2 y_6+y_1^2 y_2 y_3 y_4^3 y_5^3 y_6+2
   y_1^3 y_2 y_3 y_4^3 y_5^3 y_6+y_1^4 y_2 y_3 y_4^3 y_5^3 y_6+y_1^3 y_2^2 y_3 y_4^3 y_5^3 y_6+y_1^4
   y_2^2 y_3 y_4^3 y_5^3 y_6 \\ & & +y_1^2 y_2 y_3 y_4^4 y_5^3 y_6+2 y_1^3 y_2 y_3 y_4^4 y_5^3 y_6+y_1^4 y_2 y_3
   y_4^4 y_5^3 y_6+2 y_1^3 y_2^2 y_3 y_4^4 y_5^3 y_6+2 y_1^4 y_2^2 y_3 y_4^4 y_5^3 y_6+y_1^3 y_2^2 y_3
   y_4^5 y_5^3 y_6 \\ & & +y_1^4 y_2^2 y_3 y_4^5 y_5^3 y_6+y_4^3 y_5^2 y_6^2+4 y_1 y_4^3 y_5^2 y_6^2+6 y_1^2
   y_4^3 y_5^2 y_6^2+4 y_1^3 y_4^3 y_5^2 y_6^2+y_1^4 y_4^3 y_5^2 y_6^2+3 y_1 y_2 y_4^3 y_5^2 y_6^2+9
   y_1^2 y_2 y_4^3 y_5^2 y_6^2 \\ & & +9 y_1^3 y_2 y_4^3 y_5^2 y_6^2+3 y_1^4 y_2 y_4^3 y_5^2 y_6^2+3 y_1^2 y_2^2
   y_4^3 y_5^2 y_6^2+6 y_1^3 y_2^2 y_4^3 y_5^2 y_6^2+3 y_1^4 y_2^2 y_4^3 y_5^2 y_6^2+y_1^3 y_2^3 y_4^3
   y_5^2 y_6^2+y_1^4 y_2^3 y_4^3 y_5^2 y_6^2 \\ & & +2 y_1^2 y_2 y_3 y_4^3 y_5^2 y_6^2+4 y_1^3 y_2 y_3 y_4^3
   y_5^2 y_6^2+2 y_1^4 y_2 y_3 y_4^3 y_5^2 y_6^2+y_1^2 y_2^2 y_3 y_4^3 y_5^2 y_6^2+5 y_1^3 y_2^2 y_3
   y_4^3 y_5^2 y_6^2+4 y_1^4 y_2^2 y_3 y_4^3 y_5^2 y_6^2 \\ & & +y_1^3 y_2^3 y_3 y_4^3 y_5^2 y_6^2+2 y_1^4 y_2^3
   y_3 y_4^3 y_5^2 y_6^2+y_1^4 y_2^2 y_3^2 y_4^3 y_5^2 y_6^2+y_1^4 y_2^3 y_3^2 y_4^3 y_5^2 y_6^2+2 y_1
   y_2 y_4^4 y_5^2 y_6^2+6 y_1^2 y_2 y_4^4 y_5^2 y_6^2+6 y_1^3 y_2 y_4^4 y_5^2 y_6^2 \\ & & +2 y_1^4 y_2 y_4^4
   y_5^2 y_6^2+4 y_1^2 y_2^2 y_4^4 y_5^2 y_6^2+8 y_1^3 y_2^2 y_4^4 y_5^2 y_6^2+4 y_1^4 y_2^2 y_4^4 y_5^2
   y_6^2+2 y_1^3 y_2^3 y_4^4 y_5^2 y_6^2+2 y_1^4 y_2^3 y_4^4 y_5^2 y_6^2+2 y_1^2 y_2 y_3 y_4^4 y_5^2
   y_6^2 \\ & & +4 y_1^3 y_2 y_3 y_4^4 y_5^2 y_6^2+2 y_1^4 y_2 y_3 y_4^4 y_5^2 y_6^2+2 y_1^2 y_2^2 y_3 y_4^4
   y_5^2 y_6^2+10 y_1^3 y_2^2 y_3 y_4^4 y_5^2 y_6^2+8 y_1^4 y_2^2 y_3 y_4^4 y_5^2 y_6^2+3 y_1^3 y_2^3
   y_3 y_4^4 y_5^2 y_6^2 \\ & & +6 y_1^4 y_2^3 y_3 y_4^4 y_5^2 y_6^2+3 y_1^4 y_2^2 y_3^2 y_4^4 y_5^2 y_6^2+4
   y_1^4 y_2^3 y_3^2 y_4^4 y_5^2 y_6^2+y_1^2 y_2^2 y_4^5 y_5^2 y_6^2+2 y_1^3 y_2^2 y_4^5 y_5^2
   y_6^2+y_1^4 y_2^2 y_4^5 y_5^2 y_6^2+y_1^3 y_2^3 y_4^5 y_5^2 y_6^2 \\ & & +y_1^4 y_2^3 y_4^5 y_5^2 y_6^2+y_1^2
   y_2^2 y_3 y_4^5 y_5^2 y_6^2+5 y_1^3 y_2^2 y_3 y_4^5 y_5^2 y_6^2+4 y_1^4 y_2^2 y_3 y_4^5 y_5^2 y_6^2+3
   y_1^3 y_2^3 y_3 y_4^5 y_5^2 y_6^2+6 y_1^4 y_2^3 y_3 y_4^5 y_5^2 y_6^2 \\ & & +3 y_1^4 y_2^2 y_3^2 y_4^5 y_5^2
   y_6^2+6 y_1^4 y_2^3 y_3^2 y_4^5 y_5^2 y_6^2+y_1^3 y_2^3 y_3 y_4^6 y_5^2 y_6^2+2 y_1^4 y_2^3 y_3 y_4^6
   y_5^2 y_6^2+y_1^4 y_2^2 y_3^2 y_4^6 y_5^2 y_6^2+4 y_1^4 y_2^3 y_3^2 y_4^6 y_5^2 y_6^2 \\ & & +y_1^4 y_2^3
   y_3^2 y_4^7 y_5^2 y_6^2+y_1^2 y_2 y_3 y_4^4 y_5^3 y_6^2+3 y_1^3 y_2 y_3 y_4^4 y_5^3 y_6^2+3 y_1^4 y_2
   y_3 y_4^4 y_5^3 y_6^2+y_1^5 y_2 y_3 y_4^4 y_5^3 y_6^2+y_1^2 y_2^2 y_3 y_4^4 y_5^3 y_6^2 \\ & & +4 y_1^3 y_2^2
   y_3 y_4^4 y_5^3 y_6^2+5 y_1^4 y_2^2 y_3 y_4^4 y_5^3 y_6^2+2 y_1^5 y_2^2 y_3 y_4^4 y_5^3 y_6^2+y_1^3
   y_2^3 y_3 y_4^4 y_5^3 y_6^2+2 y_1^4 y_2^3 y_3 y_4^4 y_5^3 y_6^2+y_1^5 y_2^3 y_3 y_4^4 y_5^3
   y_6^2 \\ & & +y_1^4 y_2^2 y_3^2 y_4^4 y_5^3 y_6^2+y_1^5 y_2^2 y_3^2 y_4^4 y_5^3 y_6^2+2 y_1^4 y_2^3 y_3^2
   y_4^4 y_5^3 y_6^2+y_1^5 y_2^3 y_3^2 y_4^4 y_5^3 y_6^2+y_1^2 y_2^2 y_3 y_4^5 y_5^3 y_6^2+4 y_1^3 y_2^2
   y_3 y_4^5 y_5^3 y_6^2 \\ & & +5 y_1^4 y_2^2 y_3 y_4^5 y_5^3 y_6^2+2 y_1^5 y_2^2 y_3 y_4^5 y_5^3 y_6^2+2 y_1^3
   y_2^3 y_3 y_4^5 y_5^3 y_6^2+4 y_1^4 y_2^3 y_3 y_4^5 y_5^3 y_6^2+2 y_1^5 y_2^3 y_3 y_4^5 y_5^3 y_6^2+2
   y_1^4 y_2^2 y_3^2 y_4^5 y_5^3 y_6^2 \\ & & +2 y_1^5 y_2^2 y_3^2 y_4^5 y_5^3 y_6^2+6 y_1^4 y_2^3 y_3^2 y_4^5
   y_5^3 y_6^2+3 y_1^5 y_2^3 y_3^2 y_4^5 y_5^3 y_6^2+y_1^3 y_2^3 y_3 y_4^6 y_5^3 y_6^2+2 y_1^4 y_2^3 y_3
   y_4^6 y_5^3 y_6^2+y_1^5 y_2^3 y_3 y_4^6 y_5^3 y_6^2 \\ & & +y_1^4 y_2^2 y_3^2 y_4^6 y_5^3 y_6^2+y_1^5 y_2^2
   y_3^2 y_4^6 y_5^3 y_6^2+6 y_1^4 y_2^3 y_3^2 y_4^6 y_5^3 y_6^2+3 y_1^5 y_2^3 y_3^2 y_4^6 y_5^3 y_6^2+2
   y_1^4 y_2^3 y_3^2 y_4^7 y_5^3 y_6^2+y_1^5 y_2^3 y_3^2 y_4^7 y_5^3 y_6^2 \\ & & +y_1^4 y_2^3 y_3^2 y_4^5 y_5^4
   y_6^2+y_1^5 y_2^3 y_3^2 y_4^5 y_5^4 y_6^2+2 y_1^4 y_2^3 y_3^2 y_4^6 y_5^4 y_6^2+2 y_1^5 y_2^3 y_3^2
   y_4^6 y_5^4 y_6^2+y_1^4 y_2^3 y_3^2 y_4^7 y_5^4 y_6^2+y_1^5 y_2^3 y_3^2 y_4^7 y_5^4 y_6^2 \\ & & +y_1^2 y_2^2
   y_3 y_4^5 y_5^3 y_6^3+3 y_1^3 y_2^2 y_3 y_4^5 y_5^3 y_6^3+3 y_1^4 y_2^2 y_3 y_4^5 y_5^3 y_6^3+y_1^5
   y_2^2 y_3 y_4^5 y_5^3 y_6^3+2 y_1^3 y_2^3 y_3 y_4^5 y_5^3 y_6^3+4 y_1^4 y_2^3 y_3 y_4^5 y_5^3 y_6^3 \\ & & +2
   y_1^5 y_2^3 y_3 y_4^5 y_5^3 y_6^3+y_1^4 y_2^4 y_3 y_4^5 y_5^3 y_6^3+y_1^5 y_2^4 y_3 y_4^5 y_5^3
   y_6^3+y_1^4 y_2^3 y_3^2 y_4^5 y_5^3 y_6^3+y_1^5 y_2^3 y_3^2 y_4^5 y_5^3 y_6^3+y_1^5 y_2^4 y_3^2 y_4^5
   y_5^3 y_6^3 \\ & & +2 y_1^3 y_2^3 y_3 y_4^6 y_5^3 y_6^3+4 y_1^4 y_2^3 y_3 y_4^6 y_5^3 y_6^3+2 y_1^5 y_2^3 y_3
   y_4^6 y_5^3 y_6^3+2 y_1^4 y_2^4 y_3 y_4^6 y_5^3 y_6^3+2 y_1^5 y_2^4 y_3 y_4^6 y_5^3 y_6^3+2 y_1^4
   y_2^3 y_3^2 y_4^6 y_5^3 y_6^3 \\ & & +2 y_1^5 y_2^3 y_3^2 y_4^6 y_5^3 y_6^3+3 y_1^5 y_2^4 y_3^2 y_4^6 y_5^3
   y_6^3+y_1^4 y_2^4 y_3 y_4^7 y_5^3 y_6^3+y_1^5 y_2^4 y_3 y_4^7 y_5^3 y_6^3+y_1^4 y_2^3 y_3^2 y_4^7
   y_5^3 y_6^3+y_1^5 y_2^3 y_3^2 y_4^7 y_5^3 y_6^3 \\ & & +3 y_1^5 y_2^4 y_3^2 y_4^7 y_5^3 y_6^3+y_1^5 y_2^4
   y_3^2 y_4^8 y_5^3 y_6^3+y_1^4 y_2^3 y_3^2 y_4^6 y_5^4 y_6^3+2 y_1^5 y_2^3 y_3^2 y_4^6 y_5^4
   y_6^3+y_1^6 y_2^3 y_3^2 y_4^6 y_5^4 y_6^3+y_1^5 y_2^4 y_3^2 y_4^6 y_5^4 y_6^3 \\ & & +y_1^6 y_2^4 y_3^2 y_4^6
   y_5^4 y_6^3+y_1^4 y_2^3 y_3^2 y_4^7 y_5^4 y_6^3+2 y_1^5 y_2^3 y_3^2 y_4^7 y_5^4 y_6^3+y_1^6 y_2^3
   y_3^2 y_4^7 y_5^4 y_6^3+2 y_1^5 y_2^4 y_3^2 y_4^7 y_5^4 y_6^3+2 y_1^6 y_2^4 y_3^2 y_4^7 y_5^4
   y_6^3 \\ & & +y_1^5 y_2^4 y_3^2 y_4^8 y_5^4 y_6^3+y_1^6 y_2^4 y_3^2 y_4^8 y_5^4 y_6^3

\end{array}
\eeq
}

\newpage


\end{document}

%% file: texpref.tex
\newcommand{\eref}[1]{(\ref{#1})}

\newcommand{\fref}[1]{Figure~\ref{#1}}
\newcommand{\cref}[1]{Chapter~\ref{#1}}
\newcommand{\beq}{\begin{equation}}
\newcommand{\eeq}{\end{equation}}
\newcommand{\ba}{\begin{array}}
\newcommand{\ea}{\end{array}}
\newcommand{\bcenter}{\begin{center}}
\newcommand{\ecenter}{\end{center}}

\def\IB{\relax\hbox{$\inbar\kern-.3em{\rm B}$}}
\def\IC{\relax\hbox{$\inbar\kern-.3em{\rm C}$}}
\def\ID{\relax\hbox{$\inbar\kern-.3em{\rm D}$}}
\def\IE{\relax\hbox{$\inbar\kern-.3em{\rm E}$}}
\def\IF{\relax\hbox{$\inbar\kern-.3em{\rm F}$}}
\def\IG{\relax\hbox{$\inbar\kern-.3em{\rm G}$}}
\def\IGa{\relax\hbox{${\rm I}\kern-.18em\Gamma$}}
\def\IH{\relax{\rm I\kern-.18em H}}
\def\IK{\relax{\rm I\kern-.18em K}}
\def\IL{\relax{\rm I\kern-.18em L}}
\def\IP{\relax{\rm I\kern-.18em P}}
\def\IR{\relax{\rm I\kern-.18em R}}
\def\IZ{\relax\ifmmode\mathchoice
{\hbox{\cmss Z\kern-.4em Z}}{\hbox{\cmss Z\kern-.4em Z}}
{\lower.9pt\hbox{\cmsss Z\kern-.4em Z}}
{\lower1.2pt\hbox{\cmsss Z\kern-.4em Z}}\else{\cmss Z\kern-.4em Z}\fi}
\def\II{\relax{\rm I\kern-.18em I}}

\def\sCC{{\kern 0.27em\vrule height1.45ex width0.03em depth0em
          \kern-0.30em\rm C}}
\def\C{{\mathchoice
  {\sCC}
  {\sCC}
  {\kern 0.225em \vrule height1.05ex width0.025em depth0em \kern-0.25em \rm C}
  {\kern 0.180em \vrule height0.78ex width0.02em depth0em \kern-0.2em \rm C}
        }}
\def\sHH{{\rm I\kern-.16em{}H}}
\def\sNN{{\rm I\kern-.16em{}N}}
\def\N{{\mathchoice
  {\sNN}
  {\sNN}
  {\rm I\kern-.12em{}N}
  {\rm I\kern-.10em{}N} }}
\def\sPP{{\rm I\kern-.16em{}P}}
\def\P{{\mathchoice
  {\sPP}
  {\sPP}
  {\rm I\kern-.12em{}P}
  {\rm I\kern-.10em{}P} }}
\def\sQQ{{\kern 0.27em \vrule height1.45ex width0.03em depth0em
          \kern-0.30em \rm Q}}
\def\Q{{\mathchoice
        {\sQQ}
        {\sQQ}
  {\kern 0.225em \vrule height1.05ex width0.025em depth0em \kern-0.25em \rm Q}
  {\kern 0.180em \vrule height0.78ex width0.020em depth0em \kern-0.20em \rm Q}
        }}
\def\sRR{{\rm I\kern-0.16em{}R}}
\def\R{{\mathchoice
  {\sRR}
  {\sRR}
  {\rm I\kern-0.12em{}R}
  {\rm I\kern-0.10em{}R} }}
\def\sZZ{{\rm Z\kern-0.32em{}Z}}
\def\Z{{\mathchoice
  {\sZZ}
  {\sZZ} 
  {\rm Z\kern-0.3em{}Z}     
  {\rm Z\kern-0.25em{}Z} }}  
\def\ZZZ{{\rm Z\kern-0.24em{}Z}}
\def\sII{{\rm I\kern-0.16em{}I}}
\def\I{{\mathchoice
  {\sII}
  {\sII}
  {\rm I\kern-0.12em{}I}
  {\rm I\kern-0.10em{}I} }}

\def\Hom{{\rm Hom}}

\def\inbar{\,\vrule height1.5ex width.4pt depth0pt}
\font\cmss=cmss10 \font\cmsss=cmss10 at 7pt

\def\smiley{\hbox{\large$\bigcirc$\hspace{-0.80em}\raise.2ex
\hbox{$\cdot\cdot$}\kern-.61em\lower.2ex\hbox{\scriptsize$\smile$}}\ }
\def\frowny{\hbox{\large$\bigcirc$\hspace{-0.80em}\raise.2ex
\hbox{$\cdot\cdot$}\kern-.635em\lower.2ex\hbox{\scriptsize$\frown$}}\ }

\def\I{{\rlap{1} \hskip 1.6pt \hbox{1}}}

\makeatletter
\let\hangafter\@hangfrom
\makeatother